\documentclass[%
 reprint,
  superscriptaddress,
 amsmath,amssymb,amsfonts,
 aps,
 prl,
 floatfix,
 longbibliography
]{revtex4-2}

\usepackage[utf8]{inputenc}
\usepackage[pdftex]{graphicx} \graphicspath{{}}
\usepackage{float} \usepackage{color}
\usepackage[pdftex,colorlinks=true]{hyperref}
\usepackage{subfigure}
\hypersetup{
  colorlinks=true,
  linkcolor=blue,
  citecolor = blue,
  urlcolor=blue,
}

\usepackage{mathtools}

\DeclarePairedDelimiterX\braket[2]{\langle}{\rangle}{#1 \delimsize\vert #2}
\DeclarePairedDelimiterX\expval[3]{\langle}{\rangle}{#1 \delimsize\vert #2  \delimsize\vert #3}
\DeclarePairedDelimiterX\proj[2]{\delimsize\vert#1\rangle}{\langle#2\delimsize\vert}{ }
\usepackage{siunitx}

\begin{document}

\title{Relaxation in dipolar spin ladders: from pair production to false-vacuum decay}
\author{Gustavo A. Dom\'inguez-Castro}
\affiliation{Institut f\"ur Theoretische Physik, Leibniz Universit\"at Hannover, Appelstr. 2, D-30167 Hannover, Germany}
\author{Thomas Bilitewski}
\affiliation{Department of Physics, Oklahoma State University, Stillwater, Oklahoma 74078, USA}
\author{David Wellnitz}
\affiliation{JILA, National Institute of Standards and Technology and Department of Physics, University of Colorado, Boulder, CO, 80309, USA}
\affiliation{Center for Theory of Quantum Matter, University of Colorado, Boulder, CO, 80309, USA}
\author{Ana Maria Rey}
\affiliation{JILA, National Institute of Standards and Technology and Department of Physics, University of Colorado, Boulder, CO, 80309, USA}
\affiliation{Center for Theory of Quantum Matter, University of Colorado, Boulder, CO, 80309, USA}
\author{Luis Santos}
\affiliation{Institut f\"ur Theoretische Physik, Leibniz Universit\"at Hannover, Appelstr. 2, D-30167 Hannover, Germany}

\date{\today}

\begin{abstract}
Ultracold dipolar particles pinned in optical lattices or tweezers provide an excellent platform for the study of the intriguing equilibration dynamics of spin models with dipolar exchange.
Starting with an initial state in which spins of opposite orientation are prepared in each of the legs of a ladder lattice, we show that spin relaxation displays an unexpected dependence on inter-leg distance and dipole orientation. This dependence, stemming from the interplay between intra- and inter-leg interactions, results in three distinct relaxation regimes: (i) ergodic, characterized by the fast relaxation towards equilibrium of correlated pairs of excitations generated at exponentially fast rates from the initial state; (ii) metastable, in which the state is quasi-localized in the initial state and only decays in exceedingly long timescales, resembling false vacuum decay; and, surprisingly, (iii) partially-relaxed, with coexisting fast partial relaxation and partial quasi-localization. Realizing this intriguing dynamics is at hand of current state-of-the-art experiments in dipolar gases.
\end{abstract}
\pacs{}

\maketitle




Quantum simulators using ultracold gases in optical potentials have dramatically improved our understanding of many-body quantum dynamics, as highlighted by recent experiments on many-body localization~\cite{schreiber2015observation,PhysRevLett.114.083002,smith2016many}, 
quantum scars~\cite{bernien2017probing}, or Hilbert-space fragmentation~\cite{scherg2021observing, morong2021observation}. 
Up to recently, nearest-neighbor spin interactions only resulted from relatively weak super-exchange processes~\cite{trotzky2008time,mazurenko2017antiferromagnet,brown2015two,PhysRevLett.115.260401,hart2015observation,PhysRevLett.118.170401,brown2017spin,PhysRevLett.124.043204,nichols2019spin,PhysRevX.11.011039,hirthe2023magnetically}. This is changing due to rapid progresses in the realization of dipole-mediated spin interactions 
in various physical systems, including magnetic atoms~\cite{dePaz2013nonequilibrium, chomaz2022dipolar,PhysRevLett.129.023401,PhysRevLett.125.143401,lepoutre2019out,su2023dipolar,PhysRevResearch.2.023050}, 
Rydberg atoms~\cite{browaeys2020many,scholl2022microwave,chew2022ultrafast,signoles2021glassy},
and polar molecules~\cite{yan2013observation,Bohn_Science_357_2017,moses2017new}. In parallel, the development of optical tweezers~\cite{NatPhysKaufman2021,bao2022dipolar,holland2022ondemand,zhang2022optical} and quantum gas microscopes for polar molecules~\cite{christakis2022probing} opens fascinating perspectives for unveiling the intriguing dynamics of dipolar spin models resulting from the long-range anisotropic dipole-dipole interactions. 


Exploring non-equilibrium dynamics in dipolar spin models is further facilitated by the capability to controllably prepare layered arrays with non-trivial initial spin distributions \cite{Atala2014, Ye2019, Gall2021, Hirthe2023, wienand2023emergence}. For polar molecules this was demonstrated in recent experiments where each layer was initialized in opposite spin states~(encoded in rotational levels)~\cite{Tobias_Science_375_2022}. 
This case is particularly interesting since dipolar exchange results in the correlated creation of 
spin excitations in each layer~\cite{bilitewski2023manipulating, bilitewski2023momentumselective}, resembling pair creation from vacuum fluctuations, or parametric amplification and two-mode squeezing in quantum optics~\cite{agarwal2013quantum,PhysRevA.97.063611}. 


\begin{figure}[h!]
\centering
\includegraphics[width=\columnwidth]{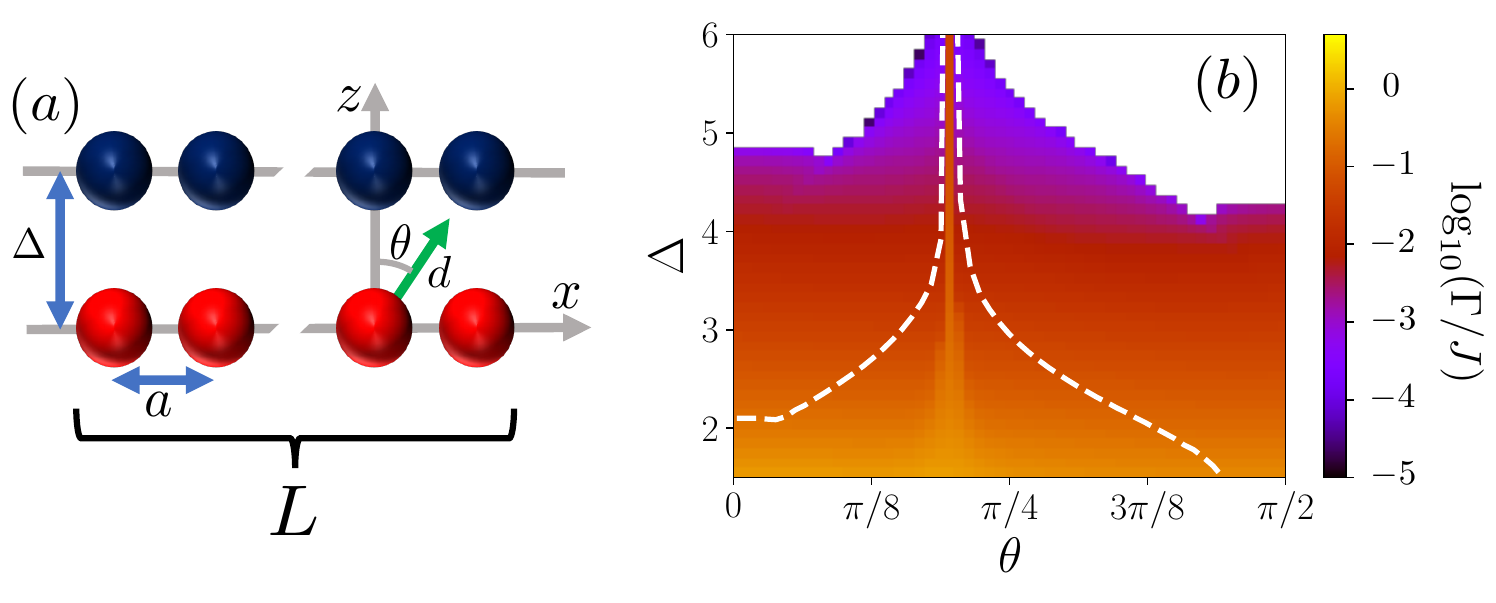}
\includegraphics[width=\columnwidth]{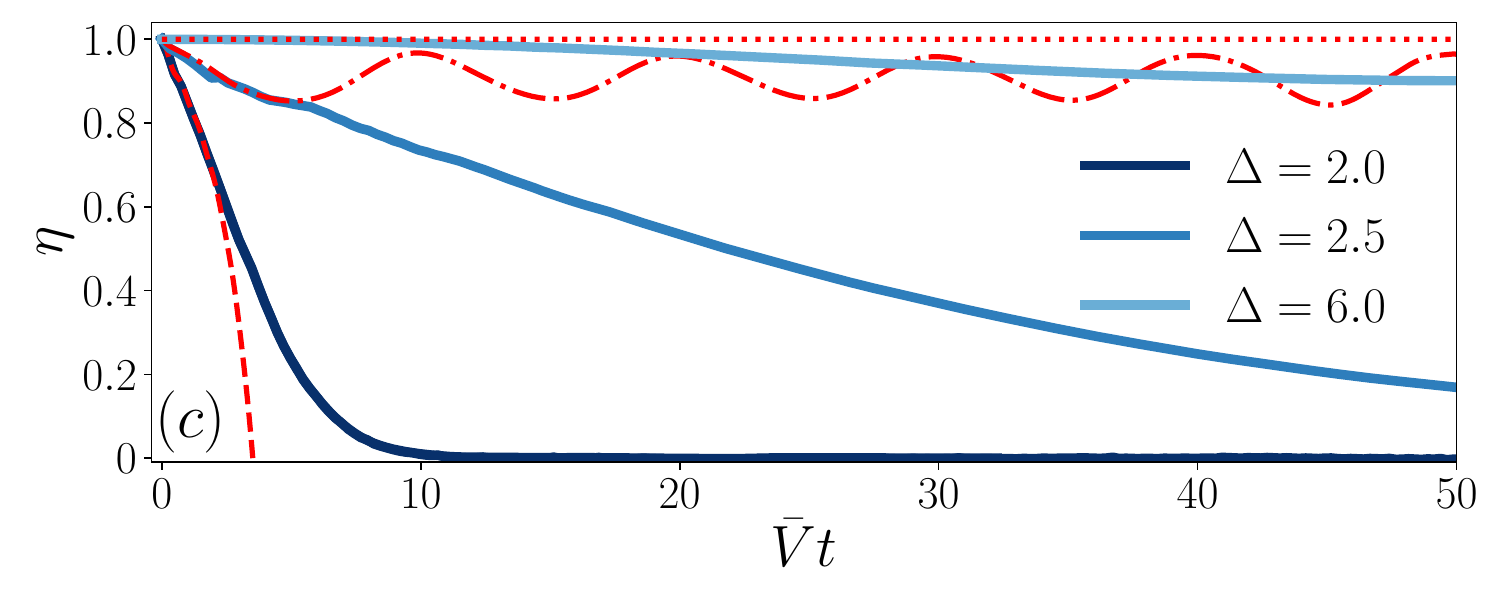}
\caption{(a) Sketch of the ladder model. Spins in each leg are initially prepared with opposite orientations~(red and blue balls). (b) Instability rate $\Gamma$ for a ladder of $L=1001$ rungs, as a function of the orientation $\theta$ and the separation $\Delta$. The colored region indicates where at least one momentum mode is unstable. The dashed curve is the effective instability threshold for $L=11$. (c) Imbalance evolution for $\theta=0$ and $\Delta=2$, $2.5$ and $6$~($a=1$); dashed, dash-dotted, and dotted red curves depict the Bogoliubov predictions for each case, respectively.
\label{fig:1}}
\end{figure}



Although in contrast to integrable~\cite{Rigol2007} and many-body localized systems~\cite{Abanin2019} non-integrable ones are expected to eventually thermalize~\cite{Dalessio2016}, they may do so after a possibly long transient prethermalization stage, which has been at the focus of major attention~\cite{Dalessio2016, Mallayya2019, Bastianello2021}. %
Long-lived metastable states also appear in so-called false vacuum decay, a phenomena first investigated in the context of quantum field theories with applications in cosmology and theory of fundamental interactions~\cite{coleman1977}, which has been predicted in quantum spin chains \cite{Lagnese2021,PRXQuantum.3.020316} and has been recently observed in ferromagnetic superfluids~\cite{Zenesini2024}. Ladder lattices provide a convenient system to study relaxation dynamics~\cite{Atala2014, Steinigweg2014, Ye2019, Rakovszky2022, wienand2023emergence}, as they are theoretically tractable, while still displaying interesting physics. 

In this Letter, we show that experimentally-feasible ladders of spin-1/2 pinned dipoles~(Fig.~\ref{fig:1}~(a)) present a highly non-trivial relaxation dynamics, characterized by an anomalously long-lived prethermal stage.  
We illustrate this for the particular experimentally relevant case in which the spins in each leg are initialized in opposite orientations, which may be subsequently admixed by inter-chain dipolar exchanges. We focus on whether and how the initial pattern relaxes towards an equilibrium state characterized by an average zero magnetization in both legs. It might be expected that relaxation is fastest when the inter-leg dipolar exchange is maximal, i.e. for dipoles oriented perpendicular to the ladder axis~($\theta=0$ in Fig.~\ref{fig:1}~(a)), and that increasing the inter-chain separation $\Delta$ only trivially enlarges the time scale of relaxation, without modifying its qualitative nature.

Interestingly, none of these a priori reasonable expectations are correct, illustrating the highly non-trivial nature of the dynamics of dipolar spins.  We show that spin relaxation is maximally favored at the dipole orientation for which there is no spin exchange along the legs. Moreover, increasing $\Delta$ does not only enlarge the relaxation time scale, but changes qualitatively the long-time evolution, leading to three distinct relaxation regimes. The ergodic regime, at low $\Delta$, is characterized by an exponential creation of correlated pairs of spin excitations at short times, followed by a fast relaxation towards equilibrium. At large $\Delta$, the system is in a highly metastable, quasi-localized, regime, remaining in the initial state and only relaxing in an exceedingly long timescale resembling false vacuum decay \cite{coleman1977,Lagnese2021,PRXQuantum.3.020316,Zenesini2024}. 
Remarkably, at intermediate $\Delta$, the system is in a partially-relaxed regime, with 
coexisting partial equilibration and very long-lived partial quasi-localization. Moreover, the three relaxation regimes present as well a markedly different evolution of the entanglement entropy~\cite{SM}.
The observation of these dynamics is well within reach of 
present experiments.



\paragraph{Model.--} We consider 
two parallel spin-$1/2$ chains of $L$ sites, denoted as A and B. The chains are characterized by a lattice spacing $a$, set to $a=1$ below, and are separated by a distance $\Delta$~(Fig. \ref{fig:1}~(a)). We assume $\Delta>a$. Due to resonant electric dipole-dipole interactions, at zero electric field the spins undergo both intra-leg and inter-leg exchange interactions. The system is then described by the Hamiltonian:
\begin{eqnarray}
\hat H &=& \sum_{\alpha=A,B} \sum_{i>j}V_{ij}^{\alpha\alpha}(\hat{s}_{i\alpha}^{+}\hat{s}_{j\alpha}^{-}+\hat{s}_{i\alpha}^{-}\hat{s}_{j\alpha}^{+}) \nonumber \\
&+& \sum_{i,j} V_{ij}^{AB}( \hat{s}_{iA}^{+}\hat{s}_{jB}^{-}+\hat{s}_{iA}^{-}\hat{s}_{jB}^{+}),
\label{eq:H}
\end{eqnarray}
with $\hat{s}_{i}^{\alpha}=\hat{\sigma}_{i}^{\alpha}/2$ the spin operators in terms of the Pauli matrices $\hat{\sigma}_{i}^{x,y,z}$ that act on the spin at site $i$. We focus our attention below on exchange couplings of the form
\begin{equation}
V_{ij}^{\alpha\alpha'} = \frac{J}{|\mathbf{r}_{i}^{\alpha}-\mathbf{r}_{j}^{\alpha'}|^{3}}\left(1-\frac{3[\hat{e}_{d}\cdot(\mathbf{r}_{i}^{\alpha}-\mathbf{r}_{j}^{\alpha'})]^{2}}{|\mathbf{r}_{i}^{\alpha}-\mathbf{r}_{j}^{\alpha'}|^{2}} \right),
\label{Eq2}
\end{equation}
where $J$ is the spin-exchange rate~(proportional to the dipole moment squared),  $\hat{e}_{d}=\hat{e}_{z}\cos\theta+\hat{e}_{x}\sin\theta$ characterizes the dipole orientation, fixed by an external magnetic field, and $\mathbf{r}_{i}^{\alpha}$ is the position of site $i$ of chain $\alpha$. Motivated by recent works on polar molecules in bilayers~\cite{Tobias_Science_375_2022}, we consider all spins in A~(B) initially in spin $\uparrow$~($\downarrow$). Below, we study the stability of elementary spin excitations on top of the initial pattern, which characterize short-time scales, and then address whether and how the system reaches equilibrium in longer times.



\begin{figure*}[t!]
\centering
\includegraphics[width=1.0\textwidth]{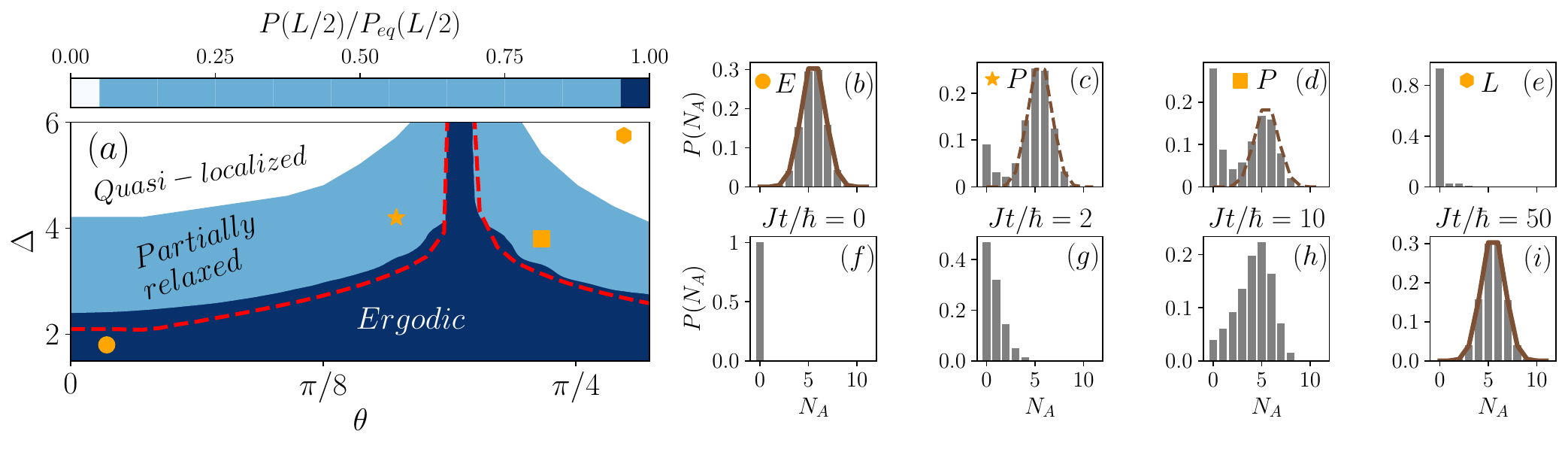}
\caption{(a) Ratio $P(L/2)/P_{eq}(L/2)$ at a time $\bar{V}t=50$ as a function of $\theta$ and $\Delta$. In the ergodic~(E) regime~(dark blue), $1-\delta<P(L/2)/P_{eq}(L/2)<1$, in the  
the partially relaxed~(P) regime~(light blue),  $\delta<P(L/2)/P_{eq}(L/2)<1-\delta$, and in the quasi-localized~(L) regime~(white), $0<P(L/2)/P_{eq}(L/2)<\delta$. We consider $\delta=0.05$, but other reasonably small values of $\delta$ do not significantly alter the results. The dashed red curve corresponds to the Bogoliubov instability threshold. (b)-(e) Distribution of spin excitations $P(N_A)$ at four points in panel (a), indicated by markers. (f)-(i) Evolution towards thermalization of $P(N_A)$ for $\Delta=2$ and $\theta=0$. Brown solid curves indicate the equilibrium distribution~\eqref{eq:P_NA}. Dashed curves depict a re-scaled Gaussian characterizing the partially-relaxed regime.  All plots are obtained using Chebyshev evolution for ladders with $11$ rungs.
\label{fig:2}}
\end{figure*}



\paragraph{Stability analysis.--}
At short times, deviations from the initial condition are best analyzed using the
Holstein-Primakoff transformation $\hat{s}_{A,i}^{z} = 1/2-\hat{a}_{i}^{\dagger}\hat{a}_{i}$, $\hat{s}_{A,i}^{+}=\hat{a}_{i}$, $\hat{s}_{B,i}^{z} = -1/2+\hat{b}_{i}^{\dagger}\hat{b}_{i}$, $\hat{s}_{B,i}^{-}=\hat{b}_{i}$, which maps the initial spin pattern to a vacuum of spin excitations~\cite{QuantumFluctuations}.  The hard-core boson operator 
$\hat a_i^\dag$~($\hat b_i^\dag$) creates a spin excitation at site $i$ in leg A~(B). For low density of spin excitations, the hard-core nature can be neglected, and the Hamiltonian may be re-expressed in quasi-momentum space as: 
\begin{equation}
\hat{H} = \sum_{k} \left [ \varepsilon_{k}(\hat{a}_{k}^{\dagger}\hat{a}_{k}+\hat{b}_{k}^{\dagger}\hat{b}_{k}) + \Omega_{k}\hat{a}_{k}^{\dagger}\hat{b}_{-k}^{\dagger}+\Omega_{k}^{*}\hat{b}_{-k}\hat{a}_{k} \right ],   
\label{eq: H_BOG}
\end{equation}
where $\hat{a}_{k}=\frac{1}{\sqrt{L}}\sum_{j}e^{-ikj}\hat{a}_{j}$ and $\hat{b}_{k}=\frac{1}{\sqrt{L}}\sum_{j}e^{-ikj}\hat{b}_{j}$. Neglecting the hard-core constraint fails at longer times.
Nevertheless, Eq.~\eqref{eq: H_BOG} provides valuable insights.
Inter-chain dipolar coupling results in the correlated creation of pairs of spin excitations of opposite momentum in each leg, with a momentum-dependent rate $\Omega_{k}=J\sum_{j}V_{0j}^{AB}e^{-ikj}$, whereas 
intra-chain interaction leads to an effective band dispersion for the motion of spin excitations along the legs,  $\varepsilon_{k}=J\sum_{j\neq 0} V_{0j}^{\alpha\alpha}e^{-ikj}
=J(1-3\sin^{2}\theta)\sum_{j\neq 0}e^{-ikj}/|j|^3$. The Hamiltonian~\eqref{eq: H_BOG} can be diagonalized by means of a Bogoliubov transformation~\cite{SM}, yielding the 
eigenenergies $\xi_{k} = \sqrt{\varepsilon_{k}^{2}-|\Omega_{k}|^2}$. Real $\xi_{k}$ lead to oscillatory dynamics $n_{k}^{A}(t) = n_{-k}^{B}(t) =  (|\Omega_{k}|/\xi_{k})^2\sin^2(\xi_{k}t)$. 
Crucially, $\xi_{k}$ may become imaginary 
for certain momenta $k_c$, resulting in the dynamical instability of the spin pattern, triggering~(at short times) the exponential growth of correlated spin excitations in both legs~\cite{bilitewski2023momentumselective}, $n_{k_{c}}^{A}(t)=n_{-k_{c}}^{B}(t) \propto(|\Omega_{k_{c}}|/|\xi_{k_{c}}|)^{2}e^{2\Gamma_{k_{c}}t}$ with $\Gamma_{k_{c}}=\text{Im} [\xi_{k_{c}}]$ the growth rate of mode $k_{c}$. 
We define the instability rate as 
$\Gamma=\max_{k_c} \Gamma_{k_c}$. Figure~\ref{fig:1}~(b) shows $\Gamma$ 
as a function of $\Delta$ and $\theta$ for two different values of $L$. 

The momentum-dependent interplay between intra-leg dispersion and inter-leg pair creation is crucial for the stability of the initial spin pattern and, as shown below, also for the long-time evolution. Bogoliubov instability for a given quasi-momentum $k$ demands $|\epsilon_k|<|\Omega_k|$. 
Note that, whereas $|\Omega_k|$ decreases with the inter-chain separation $\Delta$, the bandwidth associated with the intra-chain dispersion $\epsilon_k$ is independent of $\Delta$. As a result, except in the vicinity of the magic angle $\theta_M = \arcsin(1/\sqrt{3})$ discussed below, only $k\simeq k_0\simeq 0.46\pi$, such that $\epsilon_{k_0}=0$, contribute to the instability for large-enough $\Delta$. However, for a finite chain with $L$ sites, the quasi-momentum takes only discrete values $\{k_j\}$. Therefore for system sizes where $k_0$ cannot be reached, an effective threshold $\Delta_c(\theta,L)$ emerges~(Fig.~\ref{fig:1}~(b)), such that for  $\Delta>\Delta_c$,  $|\epsilon_{k_j}| >|\Omega_{k_j}|$ for all $k_j$, and hence the initial spin pattern is Bogoliubov stable. The situation is very different for $\theta\simeq \theta_M$, since $\epsilon_k=0$ for all $k$, and hence $\xi_k$ is imaginary for all quasi-momenta. Instability is hence strongly enhanced, being maximal for $k=0$, with $\Gamma\simeq\frac{4J}{3\Delta^2}$.



\paragraph{Magnetization imbalance and spin distribution.--}
The stability analysis cannot describe the relaxation dynamics beyond the initial stages. 
We focus at this point on the 
long-time evolution, and in particular on whether the system behaves ergodically, reaching equilibrium. To this aim, we employ 
a Chebyshev expansion~\cite{fehske2007computational, SM}, which provides a 
numerically exact evolution for arbitrarily long times~\cite{DTWA}. 
Due to its numerical complexity, we restrict to 
chains with $L=11$ rungs.
In order to characterize the relaxation dynamics, we monitor the imbalance $\eta =m_{A}-m_{B}$,
where $m_{\sigma=A,B}=\frac{1}{L}\sum_{i=1}^L \langle \hat{s}_{i\sigma}^z \rangle $ is the magnetization of chain A~(B). The imbalance is maximal, 
$\eta=1$, for the initial condition, and reaches 
$\eta=0$ when the up and down spins are evenly admixed between the two chains. Note that $\eta=1-2 n_{A}$, with $n_{A}=N_A/L$ the density of spin excitations in chain A~($N_A=\sum_i \langle \hat a_i^\dag \hat a_i \rangle$). 

A more detailed probe of the relaxation dynamics is provided by the full counting statistics $P(N_{\sigma=A,B})$ of the number of spin excitations within each chain. Due to energy conservation~($\langle \hat{H} \rangle = 0$ at any time), and the fact that spin-exchange preserves $m_A+m_B=0$, 
we expect that if ergodicity is reached, all states with 
a given number of excitations $N_A$ along chain A, and $N_B=N_A$ excitations along chain B, are equally probable, irrespective of how the excitations are spatially distributed. The probability 
that chain A has $N_{A}$ spin excitations, i.e. magnetization $m_A = 1/2-N_A/L$, would hence be~\cite{SM}:
\begin{equation}
P_{eq}(N_A) = \frac{\binom{L}{N_A}^{2}}{\binom{2L}{L}}\simeq \frac{2}{\sqrt{\pi L}} e^{-4(N_A-L/2)^{2}/L}.
\label{eq:P_NA}
\end{equation}


\begin{figure*}[t!]
\centering
\includegraphics[width=2.0\columnwidth]{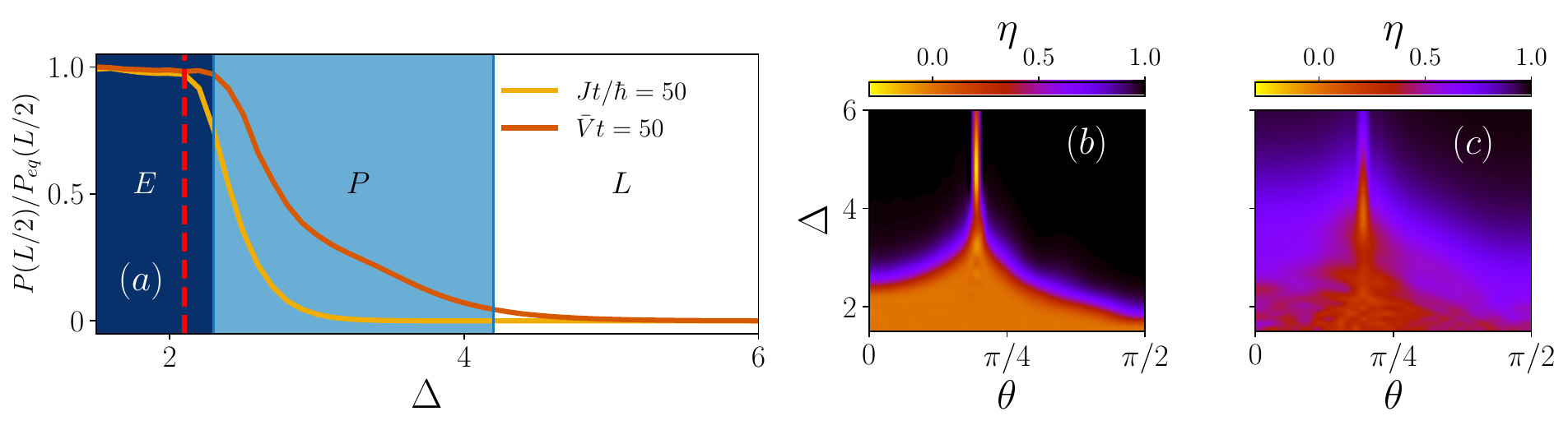}
\caption{(a) Ratio $P(L/2)/P_{eq}(L/2)$ as a function of $\Delta$ for $\theta=0$ at two different evolution times. The dashed red line corresponds to the Bogoliubov stability threshold. Colored regions indicate the ergodic (E), partially relaxed (P), and quasi-localized (L) regimes characterizing the long-time evolution.
(b)-(c) Imbalance $\eta$ after a time $Jt/\hbar=50$ as a function of $\theta$ and $\Delta$ for a fully filled lattice and for sparse~($25\%$) random filling, respectively. 
For all plots, we employ Chebyshev evolution for $11$ rungs, except in Fig.~(c), where we average over $100$ realizations of $11$ randomly placed spins in each leg of a $44$-rung ladder.
\label{fig:3}}
\end{figure*}




\paragraph{Relaxation dynamics.--} 
In order to compare the spin dynamics for different $\Delta$, we gauge out the trivial 
stretching of the time-scale of the dynamics associated to a growing $\Delta$,   by introducing the averaged inter-leg interaction
$\bar{V}=\lvert(1/L)\sum_{i,j}V_{ij}^{AB}\rvert\simeq \frac{2J}{\Delta^2}\cos^2\theta$ and considering long times, $\bar{V} t \gg 1$~\cite{footnote-angle-window}. 
Figure~\ref{fig:1}~(c) shows the evolution of the imbalance for $\theta=0$ and different $\Delta$.  For short times, numerical results are in very good agreement with the Bogoliubov prediction~(red curves). At longer times, the hard-core nature of the spin excitations, 
neglected in the  Bogoliubov analysis, 
becomes relevant. For $\Delta=2$, for which the system is Bogoliubov unstable, homogenization is quickly reached.  
For $\Delta=2.5$ and $6$, the initial pattern is Bogoliubov stable, but beyond-Bogoliubov physics eventually results in relaxation, although with 
an exceedingly long-lived memory of the initial condition.

The long-time evolution reveals three distinct relaxation regimes, illustrated in Figs.~\ref{fig:2}~(a--e) for $\bar Vt=50$.  In the ergodic regime, at low-enough $\Delta$, the system relaxes into a fully balanced distribution~($\eta=0$), and $P(N_{A})$ reaches $P_{eq}(N_A)$~(Fig.~\ref{fig:2}~(b)). Figure~\ref{fig:2}~(a) shows $P(L/2)/P_{eq}(L/2)$ as a function of $\theta$ and $\Delta$; $P(L/2)/P_{eq}(L/2)\simeq 1$ in the ergodic regime~(dark blue region). As shown in Figs.~\ref{fig:2}(f--i) for $\theta=0$ and $\Delta=2$, $P(N_A)$ rapidly approaches equilibrium, on a time of few tens of $\hbar/J$, well within experimental reach. 
As expected from Bogoliubov analysis, short-time relaxation is characterized by the exponential creation of correlated pairs of spin excitations, and hence by an exponentially decreasing $P(N_A)$~(Fig.~\ref{fig:2}(g)), and the formation of a thermo-field double state~\cite{TAKAHASHI1996,Chapman_SciPost_2019}. Eventually, further creation of excitations is arrested by their hard-core nature, and $P(N_A)$ transitions from an exponential to the Gaussian $P_{eq}(N_A)$ already at $Jt/\hbar \simeq 50$~(Fig.~\ref{fig:2}~(i)).

The long-time behavior presents similar qualitative features as the instability rate of Fig.~\ref{fig:1}~(b). The enhanced Bogoliubov instability for $\theta\simeq\theta_M$, translates into ergodic dynamics even for large $\Delta$. Moreover, as expected from the stability analysis, for large-enough $\Delta$ the system enters the quasi-localized regime, in which the spin distribution remains fully imbalanced, with $\eta\simeq 1$ and $P(N_A=0)\simeq 1$~(Fig.~\ref{fig:2}~(e)). This regime is hence characterized by $P(L/2)/P_{eq}(L/2)\simeq 0$~(white region in Fig.~\ref{fig:2}~(a)). We have checked that this remains so up to an eventual relaxation at exponentially long times, resembling the case of a metastable false vacuum~\cite{coleman1977}.

Interestingly, at intermediate $\Delta$ values, the system enters a partially-relaxed regime. $P(N_A)$ acquires a peculiar bimodal distribution~(Figs.~\ref{fig:2}(c,d)). Part reaches a Gaussian dependence around $N_A=L/2$, as expected from ergodic evolution, whereas the rest remains peaked at $N_A=0$, as in the quasi-localized regime. As a result, 
$0<P(L/2)/P_{eq}(L/2)<1$~(light blue area in Fig.~\ref{fig:2}~(a)). The bimodal distribution is maintained for very long evolution times~\cite{SM}, well over any present or foreseeable experimental lifetimes, indicating the presence of eigenstates with very different degrees of ergodicity.



\paragraph{Experimental considerations.--}
The abrupt ergodic to non-ergodic crossover 
as a function of $\Delta$
characteristic of short evolution times 
may be readily probed using polar molecules or Rydberg atoms,  
as illustrated by our results 
at $Jt/\hbar=50$ in 
Fig.~\ref{fig:3}(a), where we show $P(L/2)/P_{eq}(L/2)$ as a function of $\Delta$ for $\theta=0$~(see also Figs.~\ref{fig:2} (f--i)). Probing the partially-relaxed regime demands longer evolutions such as $\bar{V}t=50$ in Fig.~\ref{fig:3}~(a). Note that for $\Delta=3$, this corresponds to $Jt/\hbar\simeq 200$. Recent experiments on polar molecules have demonstrated rotational coherence times well over this time scale~\cite{gregory2023secondscale}.

Experiments, especially those with polar molecules in optical lattices, are characterized by a sparse lattice filling~\cite{yan2013observation, christakis2023probing}, limited to less than $25\%$. Sparse filling results in random positions of the spins, which amounts to spatial disorder of the exchange couplings. 
Compared to the clean case depicted in Fig.~\ref{fig:3}~(b), positional disorder, rather than localizing, enhances instability and  homogenization, as depicted in 
Fig.~\ref{fig:3}~(c), where we show the imbalance at $Jt/\hbar=50$ for a $25\%$ filling. In contrast, our results show that typical disorder resulting from the differential polarizability and imperfections of the lattice or the tweezer array result in an almost negligible effect~\cite{SM}. 

For the case we discussed, monitoring relaxation requires leg-resolved detection of one of the spin components. For polar molecules, this may be readily achieved by means of an electric field gradient perpendicular to the legs, which changes the energy splitting between the desired rotational states in each leg, allowing for leg-resolved spin-resolved measurements~\cite{Tobias_Science_375_2022}. %
Recent advances in single site resolution in polar molecules, magnetic and Rydberg atoms ~\cite{christakis2023probing,bornet2024enhancing,su2023dipolar} further enable the possibility to study relaxation dynamics beyond the leg-polarized initial state considered here.



\paragraph{Conclusions.--}
Dipolar spin ladders feature highly non-trivial relaxation dynamics due to the interplay between intra- and inter-leg interactions, which we have illustrated for the relevant case in which each leg is initialized in the opposite spin state. The dynamics have an intriguing dependence on the dipole orientation and the inter-chain separation, being maximally ergodic at the magic angle for which intra-chain interactions vanish. Moreover, increasing the inter-leg spacing does not only lead to the expected enlargement of time scales, but to a qualitative change in the stability of the initial pattern and in the long-term evolution, which falls into one of three possible regimes: ergodic, quasi-localized, and partially relaxed.  We have checked that other initial states, accessible in the presence of single-site resolution, such as opposite spin orientations in even and odd rungs, result in qualitatively similar relaxation regimes. The predicted dynamics may be probed in current and near-future experiments.

\acknowledgments
We acknowledge useful feedback from H. Hirzler and S. Agarwal.
L.S. and G.A.D.-C. acknowledge support of the Deutsche Forschungsgemeinschaft (DFG, German Research Foundation) under Germany’s Excellence Strategy – \ EXC-2123 QuantumFrontiers \ – \ 390837967.  A.M.R acknowledges support from the AFOSR MURI,  ARO  W911NF-24-1-0128, NSF JILA-PFC PHY-2317149, NSF QLCI-2016244, the DOE Quantum Systems Accelerator (QSA) and  NIST.

\nocite{SM}
\bibliography{Spin-Reunion}

\begin{thebibliography}{67}%
\makeatletter
\providecommand \@ifxundefined [1]{%
 \@ifx{#1\undefined}
}%
\providecommand \@ifnum [1]{%
 \ifnum #1\expandafter \@firstoftwo
 \else \expandafter \@secondoftwo
 \fi
}%
\providecommand \@ifx [1]{%
 \ifx #1\expandafter \@firstoftwo
 \else \expandafter \@secondoftwo
 \fi
}%
\providecommand \natexlab [1]{#1}%
\providecommand \enquote  [1]{``#1''}%
\providecommand \bibnamefont  [1]{#1}%
\providecommand \bibfnamefont [1]{#1}%
\providecommand \citenamefont [1]{#1}%
\providecommand \href@noop [0]{\@secondoftwo}%
\providecommand \href [0]{\begingroup \@sanitize@url \@href}%
\providecommand \@href[1]{\@@startlink{#1}\@@href}%
\providecommand \@@href[1]{\endgroup#1\@@endlink}%
\providecommand \@sanitize@url [0]{\catcode `\\12\catcode `\$12\catcode
  `\&12\catcode `\#12\catcode `\^12\catcode `\_12\catcode `\%12\relax}%
\providecommand \@@startlink[1]{}%
\providecommand \@@endlink[0]{}%
\providecommand \url  [0]{\begingroup\@sanitize@url \@url }%
\providecommand \@url [1]{\endgroup\@href {#1}{\urlprefix }}%
\providecommand \urlprefix  [0]{URL }%
\providecommand \Eprint [0]{\href }%
\providecommand \doibase [0]{https://doi.org/}%
\providecommand \selectlanguage [0]{\@gobble}%
\providecommand \bibinfo  [0]{\@secondoftwo}%
\providecommand \bibfield  [0]{\@secondoftwo}%
\providecommand \translation [1]{[#1]}%
\providecommand \BibitemOpen [0]{}%
\providecommand \bibitemStop [0]{}%
\providecommand \bibitemNoStop [0]{.\EOS\space}%
\providecommand \EOS [0]{\spacefactor3000\relax}%
\providecommand \BibitemShut  [1]{\csname bibitem#1\endcsname}%
\let\auto@bib@innerbib\@empty
\bibitem [{\citenamefont {Schreiber}\ \emph {et~al.}(2015)\citenamefont
  {Schreiber}, \citenamefont {Hodgman}, \citenamefont {Bordia}, \citenamefont
  {L{\"u}schen}, \citenamefont {Fischer}, \citenamefont {Vosk}, \citenamefont
  {Altman}, \citenamefont {Schneider},\ and\ \citenamefont
  {Bloch}}]{schreiber2015observation}%
  \BibitemOpen
  \bibfield  {author} {\bibinfo {author} {\bibfnamefont {M.}~\bibnamefont
  {Schreiber}}, \bibinfo {author} {\bibfnamefont {S.~S.}\ \bibnamefont
  {Hodgman}}, \bibinfo {author} {\bibfnamefont {P.}~\bibnamefont {Bordia}},
  \bibinfo {author} {\bibfnamefont {H.~P.}\ \bibnamefont {L{\"u}schen}},
  \bibinfo {author} {\bibfnamefont {M.~H.}\ \bibnamefont {Fischer}}, \bibinfo
  {author} {\bibfnamefont {R.}~\bibnamefont {Vosk}}, \bibinfo {author}
  {\bibfnamefont {E.}~\bibnamefont {Altman}}, \bibinfo {author} {\bibfnamefont
  {U.}~\bibnamefont {Schneider}},\ and\ \bibinfo {author} {\bibfnamefont
  {I.}~\bibnamefont {Bloch}},\ }\bibfield  {title} {\bibinfo {title}
  {Observation of many-body localization of interacting fermions in a
  quasirandom optical lattice},\ }\href
  {https://doi.org/10.1126/science.aaa7432} {\bibfield  {journal} {\bibinfo
  {journal} {Science}\ }\textbf {\bibinfo {volume} {349}},\ \bibinfo {pages}
  {842} (\bibinfo {year} {2015})}\BibitemShut {NoStop}%
\bibitem [{\citenamefont {Kondov}\ \emph {et~al.}(2015)\citenamefont {Kondov},
  \citenamefont {McGehee}, \citenamefont {Xu},\ and\ \citenamefont
  {DeMarco}}]{PhysRevLett.114.083002}%
  \BibitemOpen
  \bibfield  {author} {\bibinfo {author} {\bibfnamefont {S.~S.}\ \bibnamefont
  {Kondov}}, \bibinfo {author} {\bibfnamefont {W.~R.}\ \bibnamefont {McGehee}},
  \bibinfo {author} {\bibfnamefont {W.}~\bibnamefont {Xu}},\ and\ \bibinfo
  {author} {\bibfnamefont {B.}~\bibnamefont {DeMarco}},\ }\bibfield  {title}
  {\bibinfo {title} {Disorder-induced localization in a strongly correlated
  atomic hubbard gas},\ }\href {https://doi.org/10.1103/PhysRevLett.114.083002}
  {\bibfield  {journal} {\bibinfo  {journal} {Phys. Rev. Lett.}\ }\textbf
  {\bibinfo {volume} {114}},\ \bibinfo {pages} {083002} (\bibinfo {year}
  {2015})}\BibitemShut {NoStop}%
\bibitem [{\citenamefont {Smith}\ \emph {et~al.}(2016)\citenamefont {Smith},
  \citenamefont {Lee}, \citenamefont {Richerme}, \citenamefont {Neyenhuis},
  \citenamefont {Hess}, \citenamefont {Hauke}, \citenamefont {Heyl},
  \citenamefont {Huse},\ and\ \citenamefont {Monroe}}]{smith2016many}%
  \BibitemOpen
  \bibfield  {author} {\bibinfo {author} {\bibfnamefont {J.}~\bibnamefont
  {Smith}}, \bibinfo {author} {\bibfnamefont {A.}~\bibnamefont {Lee}}, \bibinfo
  {author} {\bibfnamefont {P.}~\bibnamefont {Richerme}}, \bibinfo {author}
  {\bibfnamefont {B.}~\bibnamefont {Neyenhuis}}, \bibinfo {author}
  {\bibfnamefont {P.~W.}\ \bibnamefont {Hess}}, \bibinfo {author}
  {\bibfnamefont {P.}~\bibnamefont {Hauke}}, \bibinfo {author} {\bibfnamefont
  {M.}~\bibnamefont {Heyl}}, \bibinfo {author} {\bibfnamefont {D.~A.}\
  \bibnamefont {Huse}},\ and\ \bibinfo {author} {\bibfnamefont
  {C.}~\bibnamefont {Monroe}},\ }\bibfield  {title} {\bibinfo {title}
  {Many-body localization in a quantum simulator with programmable random
  disorder},\ }\href {https://doi.org/10.1038/nphys3783} {\bibfield  {journal}
  {\bibinfo  {journal} {Nature Physics}\ }\textbf {\bibinfo {volume} {12}},\
  \bibinfo {pages} {907} (\bibinfo {year} {2016})}\BibitemShut {NoStop}%
\bibitem [{\citenamefont {Bernien}\ \emph {et~al.}(2017)\citenamefont
  {Bernien}, \citenamefont {Schwartz}, \citenamefont {Keesling}, \citenamefont
  {Levine}, \citenamefont {Omran}, \citenamefont {Pichler}, \citenamefont
  {Choi}, \citenamefont {Zibrov}, \citenamefont {Endres}, \citenamefont
  {Greiner} \emph {et~al.}}]{bernien2017probing}%
  \BibitemOpen
  \bibfield  {author} {\bibinfo {author} {\bibfnamefont {H.}~\bibnamefont
  {Bernien}}, \bibinfo {author} {\bibfnamefont {S.}~\bibnamefont {Schwartz}},
  \bibinfo {author} {\bibfnamefont {A.}~\bibnamefont {Keesling}}, \bibinfo
  {author} {\bibfnamefont {H.}~\bibnamefont {Levine}}, \bibinfo {author}
  {\bibfnamefont {A.}~\bibnamefont {Omran}}, \bibinfo {author} {\bibfnamefont
  {H.}~\bibnamefont {Pichler}}, \bibinfo {author} {\bibfnamefont
  {S.}~\bibnamefont {Choi}}, \bibinfo {author} {\bibfnamefont {A.~S.}\
  \bibnamefont {Zibrov}}, \bibinfo {author} {\bibfnamefont {M.}~\bibnamefont
  {Endres}}, \bibinfo {author} {\bibfnamefont {M.}~\bibnamefont {Greiner}},
  \emph {et~al.},\ }\bibfield  {title} {\bibinfo {title} {Probing many-body
  dynamics on a 51-atom quantum simulator},\ }\href
  {https://www.nature.com/articles/nature24622} {\bibfield  {journal} {\bibinfo
   {journal} {Nature}\ }\textbf {\bibinfo {volume} {551}},\ \bibinfo {pages}
  {579} (\bibinfo {year} {2017})}\BibitemShut {NoStop}%
\bibitem [{\citenamefont {Scherg}\ \emph {et~al.}(2021)\citenamefont {Scherg},
  \citenamefont {Kohlert}, \citenamefont {Sala}, \citenamefont {Pollmann},
  \citenamefont {Hebbe~Madhusudhana}, \citenamefont {Bloch},\ and\
  \citenamefont {Aidelsburger}}]{scherg2021observing}%
  \BibitemOpen
  \bibfield  {author} {\bibinfo {author} {\bibfnamefont {S.}~\bibnamefont
  {Scherg}}, \bibinfo {author} {\bibfnamefont {T.}~\bibnamefont {Kohlert}},
  \bibinfo {author} {\bibfnamefont {P.}~\bibnamefont {Sala}}, \bibinfo {author}
  {\bibfnamefont {F.}~\bibnamefont {Pollmann}}, \bibinfo {author}
  {\bibfnamefont {B.}~\bibnamefont {Hebbe~Madhusudhana}}, \bibinfo {author}
  {\bibfnamefont {I.}~\bibnamefont {Bloch}},\ and\ \bibinfo {author}
  {\bibfnamefont {M.}~\bibnamefont {Aidelsburger}},\ }\bibfield  {title}
  {\bibinfo {title} {Observing non-ergodicity due to kinetic constraints in
  tilted fermi-hubbard chains},\ }\href
  {https://www.nature.com/articles/s41467-021-24726-0} {\bibfield  {journal}
  {\bibinfo  {journal} {Nature Communications}\ }\textbf {\bibinfo {volume}
  {12}},\ \bibinfo {pages} {4490} (\bibinfo {year} {2021})}\BibitemShut
  {NoStop}%
\bibitem [{\citenamefont {Morong}\ \emph {et~al.}(2021)\citenamefont {Morong},
  \citenamefont {Liu}, \citenamefont {Becker}, \citenamefont {Collins},
  \citenamefont {Feng}, \citenamefont {Kyprianidis}, \citenamefont {Pagano},
  \citenamefont {You}, \citenamefont {Gorshkov},\ and\ \citenamefont
  {Monroe}}]{morong2021observation}%
  \BibitemOpen
  \bibfield  {author} {\bibinfo {author} {\bibfnamefont {W.}~\bibnamefont
  {Morong}}, \bibinfo {author} {\bibfnamefont {F.}~\bibnamefont {Liu}},
  \bibinfo {author} {\bibfnamefont {P.}~\bibnamefont {Becker}}, \bibinfo
  {author} {\bibfnamefont {K.}~\bibnamefont {Collins}}, \bibinfo {author}
  {\bibfnamefont {L.}~\bibnamefont {Feng}}, \bibinfo {author} {\bibfnamefont
  {A.}~\bibnamefont {Kyprianidis}}, \bibinfo {author} {\bibfnamefont
  {G.}~\bibnamefont {Pagano}}, \bibinfo {author} {\bibfnamefont
  {T.}~\bibnamefont {You}}, \bibinfo {author} {\bibfnamefont {A.}~\bibnamefont
  {Gorshkov}},\ and\ \bibinfo {author} {\bibfnamefont {C.}~\bibnamefont
  {Monroe}},\ }\bibfield  {title} {\bibinfo {title} {Observation of stark
  many-body localization without disorder},\ }\href
  {https://www.nature.com/articles/s41586-021-03988-0} {\bibfield  {journal}
  {\bibinfo  {journal} {Nature}\ }\textbf {\bibinfo {volume} {599}},\ \bibinfo
  {pages} {393} (\bibinfo {year} {2021})}\BibitemShut {NoStop}%
\bibitem [{\citenamefont {Trotzky}\ \emph {et~al.}(2008)\citenamefont
  {Trotzky}, \citenamefont {Cheinet}, \citenamefont {Folling}, \citenamefont
  {Feld}, \citenamefont {Schnorrberger}, \citenamefont {Rey}, \citenamefont
  {Polkovnikov}, \citenamefont {Demler}, \citenamefont {Lukin},\ and\
  \citenamefont {Bloch}}]{trotzky2008time}%
  \BibitemOpen
  \bibfield  {author} {\bibinfo {author} {\bibfnamefont {S.}~\bibnamefont
  {Trotzky}}, \bibinfo {author} {\bibfnamefont {P.}~\bibnamefont {Cheinet}},
  \bibinfo {author} {\bibfnamefont {S.}~\bibnamefont {Folling}}, \bibinfo
  {author} {\bibfnamefont {M.}~\bibnamefont {Feld}}, \bibinfo {author}
  {\bibfnamefont {U.}~\bibnamefont {Schnorrberger}}, \bibinfo {author}
  {\bibfnamefont {A.~M.}\ \bibnamefont {Rey}}, \bibinfo {author} {\bibfnamefont
  {A.}~\bibnamefont {Polkovnikov}}, \bibinfo {author} {\bibfnamefont {E.~A.}\
  \bibnamefont {Demler}}, \bibinfo {author} {\bibfnamefont {M.~D.}\
  \bibnamefont {Lukin}},\ and\ \bibinfo {author} {\bibfnamefont
  {I.}~\bibnamefont {Bloch}},\ }\bibfield  {title} {\bibinfo {title}
  {Time-resolved observation and control of superexchange interactions with
  ultracold atoms in optical lattices},\ }\href
  {https://doi.org/10.1126/science.1150841} {\bibfield  {journal} {\bibinfo
  {journal} {Science}\ }\textbf {\bibinfo {volume} {319}},\ \bibinfo {pages}
  {295} (\bibinfo {year} {2008})}\BibitemShut {NoStop}%
\bibitem [{\citenamefont {Mazurenko}\ \emph {et~al.}(2017)\citenamefont
  {Mazurenko}, \citenamefont {Chiu}, \citenamefont {Ji}, \citenamefont
  {Parsons}, \citenamefont {Kan{\'a}sz-Nagy}, \citenamefont {Schmidt},
  \citenamefont {Grusdt}, \citenamefont {Demler}, \citenamefont {Greif},\ and\
  \citenamefont {Greiner}}]{mazurenko2017antiferromagnet}%
  \BibitemOpen
  \bibfield  {author} {\bibinfo {author} {\bibfnamefont {A.}~\bibnamefont
  {Mazurenko}}, \bibinfo {author} {\bibfnamefont {C.~S.}\ \bibnamefont {Chiu}},
  \bibinfo {author} {\bibfnamefont {G.}~\bibnamefont {Ji}}, \bibinfo {author}
  {\bibfnamefont {M.~F.}\ \bibnamefont {Parsons}}, \bibinfo {author}
  {\bibfnamefont {M.}~\bibnamefont {Kan{\'a}sz-Nagy}}, \bibinfo {author}
  {\bibfnamefont {R.}~\bibnamefont {Schmidt}}, \bibinfo {author} {\bibfnamefont
  {F.}~\bibnamefont {Grusdt}}, \bibinfo {author} {\bibfnamefont
  {E.}~\bibnamefont {Demler}}, \bibinfo {author} {\bibfnamefont
  {D.}~\bibnamefont {Greif}},\ and\ \bibinfo {author} {\bibfnamefont
  {M.}~\bibnamefont {Greiner}},\ }\bibfield  {title} {\bibinfo {title} {A
  cold-atom fermi--hubbard antiferromagnet},\ }\href
  {https://doi.org/10.1038/nature22362} {\bibfield  {journal} {\bibinfo
  {journal} {Nature}\ }\textbf {\bibinfo {volume} {545}},\ \bibinfo {pages}
  {462} (\bibinfo {year} {2017})}\BibitemShut {NoStop}%
\bibitem [{\citenamefont {Brown}\ \emph {et~al.}(2015)\citenamefont {Brown},
  \citenamefont {Wyllie}, \citenamefont {Koller}, \citenamefont {Goldschmidt},
  \citenamefont {Foss-Feig},\ and\ \citenamefont {Porto}}]{brown2015two}%
  \BibitemOpen
  \bibfield  {author} {\bibinfo {author} {\bibfnamefont {R.}~\bibnamefont
  {Brown}}, \bibinfo {author} {\bibfnamefont {R.}~\bibnamefont {Wyllie}},
  \bibinfo {author} {\bibfnamefont {S.}~\bibnamefont {Koller}}, \bibinfo
  {author} {\bibfnamefont {E.}~\bibnamefont {Goldschmidt}}, \bibinfo {author}
  {\bibfnamefont {M.}~\bibnamefont {Foss-Feig}},\ and\ \bibinfo {author}
  {\bibfnamefont {J.}~\bibnamefont {Porto}},\ }\bibfield  {title} {\bibinfo
  {title} {Two-dimensional superexchange-mediated magnetization dynamics in an
  optical lattice},\ }\href {https://doi.org/10.1126/science.aaa1385}
  {\bibfield  {journal} {\bibinfo  {journal} {Science}\ }\textbf {\bibinfo
  {volume} {348}},\ \bibinfo {pages} {540} (\bibinfo {year}
  {2015})}\BibitemShut {NoStop}%
\bibitem [{\citenamefont {Greif}\ \emph {et~al.}(2015)\citenamefont {Greif},
  \citenamefont {Jotzu}, \citenamefont {Messer}, \citenamefont {Desbuquois},\
  and\ \citenamefont {Esslinger}}]{PhysRevLett.115.260401}%
  \BibitemOpen
  \bibfield  {author} {\bibinfo {author} {\bibfnamefont {D.}~\bibnamefont
  {Greif}}, \bibinfo {author} {\bibfnamefont {G.}~\bibnamefont {Jotzu}},
  \bibinfo {author} {\bibfnamefont {M.}~\bibnamefont {Messer}}, \bibinfo
  {author} {\bibfnamefont {R.}~\bibnamefont {Desbuquois}},\ and\ \bibinfo
  {author} {\bibfnamefont {T.}~\bibnamefont {Esslinger}},\ }\bibfield  {title}
  {\bibinfo {title} {Formation and dynamics of antiferromagnetic correlations
  in tunable optical lattices},\ }\href
  {https://doi.org/10.1103/PhysRevLett.115.260401} {\bibfield  {journal}
  {\bibinfo  {journal} {Phys. Rev. Lett.}\ }\textbf {\bibinfo {volume} {115}},\
  \bibinfo {pages} {260401} (\bibinfo {year} {2015})}\BibitemShut {NoStop}%
\bibitem [{\citenamefont {Hart}\ \emph {et~al.}(2015)\citenamefont {Hart},
  \citenamefont {Duarte}, \citenamefont {Yang}, \citenamefont {Liu},
  \citenamefont {Paiva}, \citenamefont {Khatami}, \citenamefont {Scalettar},
  \citenamefont {Trivedi}, \citenamefont {Huse},\ and\ \citenamefont
  {Hulet}}]{hart2015observation}%
  \BibitemOpen
  \bibfield  {author} {\bibinfo {author} {\bibfnamefont {R.~A.}\ \bibnamefont
  {Hart}}, \bibinfo {author} {\bibfnamefont {P.~M.}\ \bibnamefont {Duarte}},
  \bibinfo {author} {\bibfnamefont {T.-L.}\ \bibnamefont {Yang}}, \bibinfo
  {author} {\bibfnamefont {X.}~\bibnamefont {Liu}}, \bibinfo {author}
  {\bibfnamefont {T.}~\bibnamefont {Paiva}}, \bibinfo {author} {\bibfnamefont
  {E.}~\bibnamefont {Khatami}}, \bibinfo {author} {\bibfnamefont {R.~T.}\
  \bibnamefont {Scalettar}}, \bibinfo {author} {\bibfnamefont {N.}~\bibnamefont
  {Trivedi}}, \bibinfo {author} {\bibfnamefont {D.~A.}\ \bibnamefont {Huse}},\
  and\ \bibinfo {author} {\bibfnamefont {R.~G.}\ \bibnamefont {Hulet}},\
  }\bibfield  {title} {\bibinfo {title} {Observation of antiferromagnetic
  correlations in the hubbard model with ultracold atoms},\ }\href
  {https://doi.org/10.1038/nature14223} {\bibfield  {journal} {\bibinfo
  {journal} {Nature}\ }\textbf {\bibinfo {volume} {519}},\ \bibinfo {pages}
  {211} (\bibinfo {year} {2015})}\BibitemShut {NoStop}%
\bibitem [{\citenamefont {Drewes}\ \emph {et~al.}(2017)\citenamefont {Drewes},
  \citenamefont {Miller}, \citenamefont {Cocchi}, \citenamefont {Chan},
  \citenamefont {Wurz}, \citenamefont {Gall}, \citenamefont {Pertot},
  \citenamefont {Brennecke},\ and\ \citenamefont
  {K\"ohl}}]{PhysRevLett.118.170401}%
  \BibitemOpen
  \bibfield  {author} {\bibinfo {author} {\bibfnamefont {J.~H.}\ \bibnamefont
  {Drewes}}, \bibinfo {author} {\bibfnamefont {L.~A.}\ \bibnamefont {Miller}},
  \bibinfo {author} {\bibfnamefont {E.}~\bibnamefont {Cocchi}}, \bibinfo
  {author} {\bibfnamefont {C.~F.}\ \bibnamefont {Chan}}, \bibinfo {author}
  {\bibfnamefont {N.}~\bibnamefont {Wurz}}, \bibinfo {author} {\bibfnamefont
  {M.}~\bibnamefont {Gall}}, \bibinfo {author} {\bibfnamefont {D.}~\bibnamefont
  {Pertot}}, \bibinfo {author} {\bibfnamefont {F.}~\bibnamefont {Brennecke}},\
  and\ \bibinfo {author} {\bibfnamefont {M.}~\bibnamefont {K\"ohl}},\
  }\bibfield  {title} {\bibinfo {title} {Antiferromagnetic correlations in
  two-dimensional fermionic mott-insulating and metallic phases},\ }\href
  {https://doi.org/10.1103/PhysRevLett.118.170401} {\bibfield  {journal}
  {\bibinfo  {journal} {Phys. Rev. Lett.}\ }\textbf {\bibinfo {volume} {118}},\
  \bibinfo {pages} {170401} (\bibinfo {year} {2017})}\BibitemShut {NoStop}%
\bibitem [{\citenamefont {Brown}\ \emph {et~al.}(2017)\citenamefont {Brown},
  \citenamefont {Mitra}, \citenamefont {Guardado-Sanchez}, \citenamefont
  {Schau{\ss}}, \citenamefont {Kondov}, \citenamefont {Khatami}, \citenamefont
  {Paiva}, \citenamefont {Trivedi}, \citenamefont {Huse},\ and\ \citenamefont
  {Bakr}}]{brown2017spin}%
  \BibitemOpen
  \bibfield  {author} {\bibinfo {author} {\bibfnamefont {P.~T.}\ \bibnamefont
  {Brown}}, \bibinfo {author} {\bibfnamefont {D.}~\bibnamefont {Mitra}},
  \bibinfo {author} {\bibfnamefont {E.}~\bibnamefont {Guardado-Sanchez}},
  \bibinfo {author} {\bibfnamefont {P.}~\bibnamefont {Schau{\ss}}}, \bibinfo
  {author} {\bibfnamefont {S.~S.}\ \bibnamefont {Kondov}}, \bibinfo {author}
  {\bibfnamefont {E.}~\bibnamefont {Khatami}}, \bibinfo {author} {\bibfnamefont
  {T.}~\bibnamefont {Paiva}}, \bibinfo {author} {\bibfnamefont
  {N.}~\bibnamefont {Trivedi}}, \bibinfo {author} {\bibfnamefont {D.~A.}\
  \bibnamefont {Huse}},\ and\ \bibinfo {author} {\bibfnamefont {W.~S.}\
  \bibnamefont {Bakr}},\ }\bibfield  {title} {\bibinfo {title} {Spin-imbalance
  in a 2d fermi-hubbard system},\ }\href
  {https://doi.org/10.1126/science.aam7838} {\bibfield  {journal} {\bibinfo
  {journal} {Science}\ }\textbf {\bibinfo {volume} {357}},\ \bibinfo {pages}
  {1385} (\bibinfo {year} {2017})}\BibitemShut {NoStop}%
\bibitem [{\citenamefont {Dimitrova}\ \emph {et~al.}(2020)\citenamefont
  {Dimitrova}, \citenamefont {Jepsen}, \citenamefont {Buyskikh}, \citenamefont
  {Venegas-Gomez}, \citenamefont {Amato-Grill}, \citenamefont {Daley},\ and\
  \citenamefont {Ketterle}}]{PhysRevLett.124.043204}%
  \BibitemOpen
  \bibfield  {author} {\bibinfo {author} {\bibfnamefont {I.}~\bibnamefont
  {Dimitrova}}, \bibinfo {author} {\bibfnamefont {N.}~\bibnamefont {Jepsen}},
  \bibinfo {author} {\bibfnamefont {A.}~\bibnamefont {Buyskikh}}, \bibinfo
  {author} {\bibfnamefont {A.}~\bibnamefont {Venegas-Gomez}}, \bibinfo {author}
  {\bibfnamefont {J.}~\bibnamefont {Amato-Grill}}, \bibinfo {author}
  {\bibfnamefont {A.}~\bibnamefont {Daley}},\ and\ \bibinfo {author}
  {\bibfnamefont {W.}~\bibnamefont {Ketterle}},\ }\bibfield  {title} {\bibinfo
  {title} {Enhanced superexchange in a tilted mott insulator},\ }\href
  {https://doi.org/10.1103/PhysRevLett.124.043204} {\bibfield  {journal}
  {\bibinfo  {journal} {Phys. Rev. Lett.}\ }\textbf {\bibinfo {volume} {124}},\
  \bibinfo {pages} {043204} (\bibinfo {year} {2020})}\BibitemShut {NoStop}%
\bibitem [{\citenamefont {Nichols}\ \emph {et~al.}(2019)\citenamefont
  {Nichols}, \citenamefont {Cheuk}, \citenamefont {Okan}, \citenamefont
  {Hartke}, \citenamefont {Mendez}, \citenamefont {Senthil}, \citenamefont
  {Khatami}, \citenamefont {Zhang},\ and\ \citenamefont
  {Zwierlein}}]{nichols2019spin}%
  \BibitemOpen
  \bibfield  {author} {\bibinfo {author} {\bibfnamefont {M.~A.}\ \bibnamefont
  {Nichols}}, \bibinfo {author} {\bibfnamefont {L.~W.}\ \bibnamefont {Cheuk}},
  \bibinfo {author} {\bibfnamefont {M.}~\bibnamefont {Okan}}, \bibinfo {author}
  {\bibfnamefont {T.~R.}\ \bibnamefont {Hartke}}, \bibinfo {author}
  {\bibfnamefont {E.}~\bibnamefont {Mendez}}, \bibinfo {author} {\bibfnamefont
  {T.}~\bibnamefont {Senthil}}, \bibinfo {author} {\bibfnamefont
  {E.}~\bibnamefont {Khatami}}, \bibinfo {author} {\bibfnamefont
  {H.}~\bibnamefont {Zhang}},\ and\ \bibinfo {author} {\bibfnamefont {M.~W.}\
  \bibnamefont {Zwierlein}},\ }\bibfield  {title} {\bibinfo {title} {Spin
  transport in a mott insulator of ultracold fermions},\ }\href
  {https://doi.org/10.1126/science.aat4387} {\bibfield  {journal} {\bibinfo
  {journal} {Science}\ }\textbf {\bibinfo {volume} {363}},\ \bibinfo {pages}
  {383} (\bibinfo {year} {2019})}\BibitemShut {NoStop}%
\bibitem [{\citenamefont {Lu}\ \emph {et~al.}(2021)\citenamefont {Lu},
  \citenamefont {Sun}, \citenamefont {Kumar}, \citenamefont {Wang},
  \citenamefont {Gu}, \citenamefont {Zeng}, \citenamefont {Hao}, \citenamefont
  {Li}, \citenamefont {Shao}, \citenamefont {Ma}, \citenamefont {Hao},
  \citenamefont {Zhang}, \citenamefont {Mansuer}, \citenamefont {Mei},
  \citenamefont {Zhao}, \citenamefont {Liu}, \citenamefont {Deng},
  \citenamefont {Huang}, \citenamefont {Shen}, \citenamefont {Shimada},
  \citenamefont {Schwier}, \citenamefont {Liu}, \citenamefont {Liu},\ and\
  \citenamefont {Chen}}]{PhysRevX.11.011039}%
  \BibitemOpen
  \bibfield  {author} {\bibinfo {author} {\bibfnamefont {R.}~\bibnamefont
  {Lu}}, \bibinfo {author} {\bibfnamefont {H.}~\bibnamefont {Sun}}, \bibinfo
  {author} {\bibfnamefont {S.}~\bibnamefont {Kumar}}, \bibinfo {author}
  {\bibfnamefont {Y.}~\bibnamefont {Wang}}, \bibinfo {author} {\bibfnamefont
  {M.}~\bibnamefont {Gu}}, \bibinfo {author} {\bibfnamefont {M.}~\bibnamefont
  {Zeng}}, \bibinfo {author} {\bibfnamefont {Y.-J.}\ \bibnamefont {Hao}},
  \bibinfo {author} {\bibfnamefont {J.}~\bibnamefont {Li}}, \bibinfo {author}
  {\bibfnamefont {J.}~\bibnamefont {Shao}}, \bibinfo {author} {\bibfnamefont
  {X.-M.}\ \bibnamefont {Ma}}, \bibinfo {author} {\bibfnamefont
  {Z.}~\bibnamefont {Hao}}, \bibinfo {author} {\bibfnamefont {K.}~\bibnamefont
  {Zhang}}, \bibinfo {author} {\bibfnamefont {W.}~\bibnamefont {Mansuer}},
  \bibinfo {author} {\bibfnamefont {J.}~\bibnamefont {Mei}}, \bibinfo {author}
  {\bibfnamefont {Y.}~\bibnamefont {Zhao}}, \bibinfo {author} {\bibfnamefont
  {C.}~\bibnamefont {Liu}}, \bibinfo {author} {\bibfnamefont {K.}~\bibnamefont
  {Deng}}, \bibinfo {author} {\bibfnamefont {W.}~\bibnamefont {Huang}},
  \bibinfo {author} {\bibfnamefont {B.}~\bibnamefont {Shen}}, \bibinfo {author}
  {\bibfnamefont {K.}~\bibnamefont {Shimada}}, \bibinfo {author} {\bibfnamefont
  {E.~F.}\ \bibnamefont {Schwier}}, \bibinfo {author} {\bibfnamefont
  {C.}~\bibnamefont {Liu}}, \bibinfo {author} {\bibfnamefont {Q.}~\bibnamefont
  {Liu}},\ and\ \bibinfo {author} {\bibfnamefont {C.}~\bibnamefont {Chen}},\
  }\bibfield  {title} {\bibinfo {title} {Half-magnetic topological insulator
  with magnetization-induced dirac gap at a selected surface},\ }\href
  {https://doi.org/10.1103/PhysRevX.11.011039} {\bibfield  {journal} {\bibinfo
  {journal} {Phys. Rev. X}\ }\textbf {\bibinfo {volume} {11}},\ \bibinfo
  {pages} {011039} (\bibinfo {year} {2021})}\BibitemShut {NoStop}%
\bibitem [{\citenamefont {Hirthe}\ \emph
  {et~al.}(2023{\natexlab{a}})\citenamefont {Hirthe}, \citenamefont {Chalopin},
  \citenamefont {Bourgund}, \citenamefont {Bojovi{\'c}}, \citenamefont
  {Bohrdt}, \citenamefont {Demler}, \citenamefont {Grusdt}, \citenamefont
  {Bloch},\ and\ \citenamefont {Hilker}}]{hirthe2023magnetically}%
  \BibitemOpen
  \bibfield  {author} {\bibinfo {author} {\bibfnamefont {S.}~\bibnamefont
  {Hirthe}}, \bibinfo {author} {\bibfnamefont {T.}~\bibnamefont {Chalopin}},
  \bibinfo {author} {\bibfnamefont {D.}~\bibnamefont {Bourgund}}, \bibinfo
  {author} {\bibfnamefont {P.}~\bibnamefont {Bojovi{\'c}}}, \bibinfo {author}
  {\bibfnamefont {A.}~\bibnamefont {Bohrdt}}, \bibinfo {author} {\bibfnamefont
  {E.}~\bibnamefont {Demler}}, \bibinfo {author} {\bibfnamefont
  {F.}~\bibnamefont {Grusdt}}, \bibinfo {author} {\bibfnamefont
  {I.}~\bibnamefont {Bloch}},\ and\ \bibinfo {author} {\bibfnamefont {T.~A.}\
  \bibnamefont {Hilker}},\ }\bibfield  {title} {\bibinfo {title} {Magnetically
  mediated hole pairing in fermionic ladders of ultracold atoms},\ }\href
  {https://doi.org/10.1038/s41586-022-05437-y} {\bibfield  {journal} {\bibinfo
  {journal} {Nature}\ }\textbf {\bibinfo {volume} {613}},\ \bibinfo {pages}
  {463} (\bibinfo {year} {2023}{\natexlab{a}})}\BibitemShut {NoStop}%
\bibitem [{\citenamefont {de~Paz}\ \emph {et~al.}(2013)\citenamefont {de~Paz},
  \citenamefont {Sharma}, \citenamefont {Chotia}, \citenamefont {Mar\'echal},
  \citenamefont {Huckans}, \citenamefont {Pedri}, \citenamefont {Santos},
  \citenamefont {Gorceix}, \citenamefont {Vernac},\ and\ \citenamefont
  {Laburthe-Tolra}}]{dePaz2013nonequilibrium}%
  \BibitemOpen
  \bibfield  {author} {\bibinfo {author} {\bibfnamefont {A.}~\bibnamefont
  {de~Paz}}, \bibinfo {author} {\bibfnamefont {A.}~\bibnamefont {Sharma}},
  \bibinfo {author} {\bibfnamefont {A.}~\bibnamefont {Chotia}}, \bibinfo
  {author} {\bibfnamefont {E.}~\bibnamefont {Mar\'echal}}, \bibinfo {author}
  {\bibfnamefont {J.~H.}\ \bibnamefont {Huckans}}, \bibinfo {author}
  {\bibfnamefont {P.}~\bibnamefont {Pedri}}, \bibinfo {author} {\bibfnamefont
  {L.}~\bibnamefont {Santos}}, \bibinfo {author} {\bibfnamefont
  {O.}~\bibnamefont {Gorceix}}, \bibinfo {author} {\bibfnamefont
  {L.}~\bibnamefont {Vernac}},\ and\ \bibinfo {author} {\bibfnamefont
  {B.}~\bibnamefont {Laburthe-Tolra}},\ }\bibfield  {title} {\bibinfo {title}
  {Nonequilibrium quantum magnetism in a dipolar lattice gas},\ }\href
  {https://doi.org/10.1103/PhysRevLett.111.185305} {\bibfield  {journal}
  {\bibinfo  {journal} {Phys. Rev. Lett.}\ }\textbf {\bibinfo {volume} {111}},\
  \bibinfo {pages} {185305} (\bibinfo {year} {2013})}\BibitemShut {NoStop}%
\bibitem [{\citenamefont {{Chomaz}}\ \emph {et~al.}(2023)\citenamefont
  {{Chomaz}}, \citenamefont {{Ferrier-Barbut}}, \citenamefont {{Ferlaino}},
  \citenamefont {{Laburthe-Tolra}}, \citenamefont {{Lev}},\ and\ \citenamefont
  {{Pfau}}}]{chomaz2022dipolar}%
  \BibitemOpen
  \bibfield  {author} {\bibinfo {author} {\bibfnamefont {L.}~\bibnamefont
  {{Chomaz}}}, \bibinfo {author} {\bibfnamefont {I.}~\bibnamefont
  {{Ferrier-Barbut}}}, \bibinfo {author} {\bibfnamefont {F.}~\bibnamefont
  {{Ferlaino}}}, \bibinfo {author} {\bibfnamefont {B.}~\bibnamefont
  {{Laburthe-Tolra}}}, \bibinfo {author} {\bibfnamefont {B.~L.}\ \bibnamefont
  {{Lev}}},\ and\ \bibinfo {author} {\bibfnamefont {T.}~\bibnamefont
  {{Pfau}}},\ }\bibfield  {title} {\bibinfo {title} {{Dipolar physics: a review
  of experiments with magnetic quantum gases}},\ }\href
  {https://doi.org/10.1088/1361-6633/aca814} {\bibfield  {journal} {\bibinfo
  {journal} {Reports on Progress in Physics}\ }\textbf {\bibinfo {volume}
  {86}},\ \bibinfo {eid} {026401} (\bibinfo {year} {2023})}\BibitemShut
  {NoStop}%
\bibitem [{\citenamefont {Alaoui}\ \emph {et~al.}(2022)\citenamefont {Alaoui},
  \citenamefont {Zhu}, \citenamefont {Muleady}, \citenamefont {Dubosclard},
  \citenamefont {Roscilde}, \citenamefont {Rey}, \citenamefont
  {Laburthe-Tolra},\ and\ \citenamefont {Vernac}}]{PhysRevLett.129.023401}%
  \BibitemOpen
  \bibfield  {author} {\bibinfo {author} {\bibfnamefont {Y.~A.}\ \bibnamefont
  {Alaoui}}, \bibinfo {author} {\bibfnamefont {B.}~\bibnamefont {Zhu}},
  \bibinfo {author} {\bibfnamefont {S.~R.}\ \bibnamefont {Muleady}}, \bibinfo
  {author} {\bibfnamefont {W.}~\bibnamefont {Dubosclard}}, \bibinfo {author}
  {\bibfnamefont {T.}~\bibnamefont {Roscilde}}, \bibinfo {author}
  {\bibfnamefont {A.~M.}\ \bibnamefont {Rey}}, \bibinfo {author} {\bibfnamefont
  {B.}~\bibnamefont {Laburthe-Tolra}},\ and\ \bibinfo {author} {\bibfnamefont
  {L.}~\bibnamefont {Vernac}},\ }\bibfield  {title} {\bibinfo {title}
  {Measuring correlations from the collective spin fluctuations of a large
  ensemble of lattice-trapped dipolar spin-3 atoms},\ }\href
  {https://doi.org/10.1103/PhysRevLett.129.023401} {\bibfield  {journal}
  {\bibinfo  {journal} {Phys. Rev. Lett.}\ }\textbf {\bibinfo {volume} {129}},\
  \bibinfo {pages} {023401} (\bibinfo {year} {2022})}\BibitemShut {NoStop}%
\bibitem [{\citenamefont {Gabardos}\ \emph {et~al.}(2020)\citenamefont
  {Gabardos}, \citenamefont {Zhu}, \citenamefont {Lepoutre}, \citenamefont
  {Rey}, \citenamefont {Laburthe-Tolra},\ and\ \citenamefont
  {Vernac}}]{PhysRevLett.125.143401}%
  \BibitemOpen
  \bibfield  {author} {\bibinfo {author} {\bibfnamefont {L.}~\bibnamefont
  {Gabardos}}, \bibinfo {author} {\bibfnamefont {B.}~\bibnamefont {Zhu}},
  \bibinfo {author} {\bibfnamefont {S.}~\bibnamefont {Lepoutre}}, \bibinfo
  {author} {\bibfnamefont {A.~M.}\ \bibnamefont {Rey}}, \bibinfo {author}
  {\bibfnamefont {B.}~\bibnamefont {Laburthe-Tolra}},\ and\ \bibinfo {author}
  {\bibfnamefont {L.}~\bibnamefont {Vernac}},\ }\bibfield  {title} {\bibinfo
  {title} {Relaxation of the collective magnetization of a dense 3d array of
  interacting dipolar $s=3$ atoms},\ }\href
  {https://doi.org/10.1103/PhysRevLett.125.143401} {\bibfield  {journal}
  {\bibinfo  {journal} {Phys. Rev. Lett.}\ }\textbf {\bibinfo {volume} {125}},\
  \bibinfo {pages} {143401} (\bibinfo {year} {2020})}\BibitemShut {NoStop}%
\bibitem [{\citenamefont {Lepoutre}\ \emph {et~al.}(2019)\citenamefont
  {Lepoutre}, \citenamefont {Schachenmayer}, \citenamefont {Gabardos},
  \citenamefont {Zhu}, \citenamefont {Naylor}, \citenamefont {Mar{\'e}chal},
  \citenamefont {Gorceix}, \citenamefont {Rey}, \citenamefont {Vernac},\ and\
  \citenamefont {Laburthe-Tolra}}]{lepoutre2019out}%
  \BibitemOpen
  \bibfield  {author} {\bibinfo {author} {\bibfnamefont {S.}~\bibnamefont
  {Lepoutre}}, \bibinfo {author} {\bibfnamefont {J.}~\bibnamefont
  {Schachenmayer}}, \bibinfo {author} {\bibfnamefont {L.}~\bibnamefont
  {Gabardos}}, \bibinfo {author} {\bibfnamefont {B.}~\bibnamefont {Zhu}},
  \bibinfo {author} {\bibfnamefont {B.}~\bibnamefont {Naylor}}, \bibinfo
  {author} {\bibfnamefont {E.}~\bibnamefont {Mar{\'e}chal}}, \bibinfo {author}
  {\bibfnamefont {O.}~\bibnamefont {Gorceix}}, \bibinfo {author} {\bibfnamefont
  {A.~M.}\ \bibnamefont {Rey}}, \bibinfo {author} {\bibfnamefont
  {L.}~\bibnamefont {Vernac}},\ and\ \bibinfo {author} {\bibfnamefont
  {B.}~\bibnamefont {Laburthe-Tolra}},\ }\bibfield  {title} {\bibinfo {title}
  {Out-of-equilibrium quantum magnetism and thermalization in a spin-3
  many-body dipolar lattice system},\ }\href
  {https://doi.org/10.1038/s41467-019-09699-5} {\bibfield  {journal} {\bibinfo
  {journal} {Nature communications}\ }\textbf {\bibinfo {volume} {10}},\
  \bibinfo {pages} {1714} (\bibinfo {year} {2019})}\BibitemShut {NoStop}%
\bibitem [{\citenamefont {Su}\ \emph {et~al.}(2023)\citenamefont {Su},
  \citenamefont {Douglas}, \citenamefont {Szurek}, \citenamefont {Groth},
  \citenamefont {Ozturk}, \citenamefont {Krahn}, \citenamefont {Hébert},
  \citenamefont {Phelps}, \citenamefont {Ebadi}, \citenamefont {Dickerson},
  \citenamefont {Ferlaino}, \citenamefont {Marković},\ and\ \citenamefont
  {Greiner}}]{su2023dipolar}%
  \BibitemOpen
  \bibfield  {author} {\bibinfo {author} {\bibfnamefont {L.}~\bibnamefont
  {Su}}, \bibinfo {author} {\bibfnamefont {A.}~\bibnamefont {Douglas}},
  \bibinfo {author} {\bibfnamefont {M.}~\bibnamefont {Szurek}}, \bibinfo
  {author} {\bibfnamefont {R.}~\bibnamefont {Groth}}, \bibinfo {author}
  {\bibfnamefont {S.~F.}\ \bibnamefont {Ozturk}}, \bibinfo {author}
  {\bibfnamefont {A.}~\bibnamefont {Krahn}}, \bibinfo {author} {\bibfnamefont
  {A.~H.}\ \bibnamefont {Hébert}}, \bibinfo {author} {\bibfnamefont {G.~A.}\
  \bibnamefont {Phelps}}, \bibinfo {author} {\bibfnamefont {S.}~\bibnamefont
  {Ebadi}}, \bibinfo {author} {\bibfnamefont {S.}~\bibnamefont {Dickerson}},
  \bibinfo {author} {\bibfnamefont {F.}~\bibnamefont {Ferlaino}}, \bibinfo
  {author} {\bibfnamefont {O.}~\bibnamefont {Marković}},\ and\ \bibinfo
  {author} {\bibfnamefont {M.}~\bibnamefont {Greiner}},\ }\bibfield  {title}
  {\bibinfo {title} {Dipolar quantum solids emerging in a hubbard quantum
  simulator},\ }\href {https://www.nature.com/articles/s41586-023-06614-3}
  {\bibfield  {journal} {\bibinfo  {journal} {Nature}\ }\textbf {\bibinfo
  {volume} {622}},\ \bibinfo {pages} {724} (\bibinfo {year}
  {2023})}\BibitemShut {NoStop}%
\bibitem [{\citenamefont {Patscheider}\ \emph {et~al.}(2020)\citenamefont
  {Patscheider}, \citenamefont {Zhu}, \citenamefont {Chomaz}, \citenamefont
  {Petter}, \citenamefont {Baier}, \citenamefont {Rey}, \citenamefont
  {Ferlaino},\ and\ \citenamefont {Mark}}]{PhysRevResearch.2.023050}%
  \BibitemOpen
  \bibfield  {author} {\bibinfo {author} {\bibfnamefont {A.}~\bibnamefont
  {Patscheider}}, \bibinfo {author} {\bibfnamefont {B.}~\bibnamefont {Zhu}},
  \bibinfo {author} {\bibfnamefont {L.}~\bibnamefont {Chomaz}}, \bibinfo
  {author} {\bibfnamefont {D.}~\bibnamefont {Petter}}, \bibinfo {author}
  {\bibfnamefont {S.}~\bibnamefont {Baier}}, \bibinfo {author} {\bibfnamefont
  {A.-M.}\ \bibnamefont {Rey}}, \bibinfo {author} {\bibfnamefont
  {F.}~\bibnamefont {Ferlaino}},\ and\ \bibinfo {author} {\bibfnamefont
  {M.~J.}\ \bibnamefont {Mark}},\ }\bibfield  {title} {\bibinfo {title}
  {Controlling dipolar exchange interactions in a dense three-dimensional array
  of large-spin fermions},\ }\href
  {https://doi.org/10.1103/PhysRevResearch.2.023050} {\bibfield  {journal}
  {\bibinfo  {journal} {Phys. Rev. Res.}\ }\textbf {\bibinfo {volume} {2}},\
  \bibinfo {pages} {023050} (\bibinfo {year} {2020})}\BibitemShut {NoStop}%
\bibitem [{\citenamefont {Browaeys}\ and\ \citenamefont
  {Lahaye}(2020)}]{browaeys2020many}%
  \BibitemOpen
  \bibfield  {author} {\bibinfo {author} {\bibfnamefont {A.}~\bibnamefont
  {Browaeys}}\ and\ \bibinfo {author} {\bibfnamefont {T.}~\bibnamefont
  {Lahaye}},\ }\bibfield  {title} {\bibinfo {title} {{Many-body physics with
  individually controlled Rydberg atoms}},\ }\href
  {https://doi.org/10.1038/s41567-019-0733-z} {\bibfield  {journal} {\bibinfo
  {journal} {Nat. Phys.}\ }\textbf {\bibinfo {volume} {16}},\ \bibinfo {pages}
  {132} (\bibinfo {year} {2020})}\BibitemShut {NoStop}%
\bibitem [{\citenamefont {Scholl}\ \emph {et~al.}(2022)\citenamefont {Scholl},
  \citenamefont {Williams}, \citenamefont {Bornet}, \citenamefont {Wallner},
  \citenamefont {Barredo}, \citenamefont {Henriet}, \citenamefont {Signoles},
  \citenamefont {Hainaut}, \citenamefont {Franz}, \citenamefont {Geier},
  \citenamefont {Tebben}, \citenamefont {Salzinger}, \citenamefont {Z\"urn},
  \citenamefont {Lahaye}, \citenamefont {Weidem\"uller},\ and\ \citenamefont
  {Browaeys}}]{scholl2022microwave}%
  \BibitemOpen
  \bibfield  {author} {\bibinfo {author} {\bibfnamefont {P.}~\bibnamefont
  {Scholl}}, \bibinfo {author} {\bibfnamefont {H.~J.}\ \bibnamefont
  {Williams}}, \bibinfo {author} {\bibfnamefont {G.}~\bibnamefont {Bornet}},
  \bibinfo {author} {\bibfnamefont {F.}~\bibnamefont {Wallner}}, \bibinfo
  {author} {\bibfnamefont {D.}~\bibnamefont {Barredo}}, \bibinfo {author}
  {\bibfnamefont {L.}~\bibnamefont {Henriet}}, \bibinfo {author} {\bibfnamefont
  {A.}~\bibnamefont {Signoles}}, \bibinfo {author} {\bibfnamefont
  {C.}~\bibnamefont {Hainaut}}, \bibinfo {author} {\bibfnamefont
  {T.}~\bibnamefont {Franz}}, \bibinfo {author} {\bibfnamefont
  {S.}~\bibnamefont {Geier}}, \bibinfo {author} {\bibfnamefont
  {A.}~\bibnamefont {Tebben}}, \bibinfo {author} {\bibfnamefont
  {A.}~\bibnamefont {Salzinger}}, \bibinfo {author} {\bibfnamefont
  {G.}~\bibnamefont {Z\"urn}}, \bibinfo {author} {\bibfnamefont
  {T.}~\bibnamefont {Lahaye}}, \bibinfo {author} {\bibfnamefont
  {M.}~\bibnamefont {Weidem\"uller}},\ and\ \bibinfo {author} {\bibfnamefont
  {A.}~\bibnamefont {Browaeys}},\ }\bibfield  {title} {\bibinfo {title}
  {Microwave engineering of programmable $xxz$ hamiltonians in arrays of
  rydberg atoms},\ }\href {https://doi.org/10.1103/PRXQuantum.3.020303}
  {\bibfield  {journal} {\bibinfo  {journal} {PRX Quantum}\ }\textbf {\bibinfo
  {volume} {3}},\ \bibinfo {pages} {020303} (\bibinfo {year}
  {2022})}\BibitemShut {NoStop}%
\bibitem [{\citenamefont {Chew}\ \emph {et~al.}(2022)\citenamefont {Chew},
  \citenamefont {Tomita}, \citenamefont {Mahesh}, \citenamefont {Sugawa},
  \citenamefont
  {de~L{\ifmmode\acute{e}\else\'{e}\fi}s{\ifmmode\acute{e}\else\'{e}\fi}leuc},\
  and\ \citenamefont {Ohmori}}]{chew2022ultrafast}%
  \BibitemOpen
  \bibfield  {author} {\bibinfo {author} {\bibfnamefont {Y.}~\bibnamefont
  {Chew}}, \bibinfo {author} {\bibfnamefont {T.}~\bibnamefont {Tomita}},
  \bibinfo {author} {\bibfnamefont {T.~P.}\ \bibnamefont {Mahesh}}, \bibinfo
  {author} {\bibfnamefont {S.}~\bibnamefont {Sugawa}}, \bibinfo {author}
  {\bibfnamefont {S.}~\bibnamefont
  {de~L{\ifmmode\acute{e}\else\'{e}\fi}s{\ifmmode\acute{e}\else\'{e}\fi}leuc}},\
  and\ \bibinfo {author} {\bibfnamefont {K.}~\bibnamefont {Ohmori}},\
  }\bibfield  {title} {\bibinfo {title} {{Ultrafast energy exchange between two
  single Rydberg atoms on a nanosecond timescale}},\ }\href
  {https://doi.org/10.1038/s41566-022-01047-2} {\bibfield  {journal} {\bibinfo
  {journal} {Nat. Photonics}\ }\textbf {\bibinfo {volume} {16}},\ \bibinfo
  {pages} {724} (\bibinfo {year} {2022})}\BibitemShut {NoStop}%
\bibitem [{\citenamefont {Signoles}\ \emph {et~al.}(2021)\citenamefont
  {Signoles}, \citenamefont {Franz}, \citenamefont {Ferracini~Alves},
  \citenamefont {G{\ifmmode\ddot{a}\else\"{a}\fi}rttner}, \citenamefont
  {Whitlock}, \citenamefont {Z{\ifmmode\ddot{u}\else\"{u}\fi}rn},\ and\
  \citenamefont
  {Weidem{\ifmmode\ddot{u}\else\"{u}\fi}ller}}]{signoles2021glassy}%
  \BibitemOpen
  \bibfield  {author} {\bibinfo {author} {\bibfnamefont {A.}~\bibnamefont
  {Signoles}}, \bibinfo {author} {\bibfnamefont {T.}~\bibnamefont {Franz}},
  \bibinfo {author} {\bibfnamefont {R.}~\bibnamefont {Ferracini~Alves}},
  \bibinfo {author} {\bibfnamefont {M.}~\bibnamefont
  {G{\ifmmode\ddot{a}\else\"{a}\fi}rttner}}, \bibinfo {author} {\bibfnamefont
  {S.}~\bibnamefont {Whitlock}}, \bibinfo {author} {\bibfnamefont
  {G.}~\bibnamefont {Z{\ifmmode\ddot{u}\else\"{u}\fi}rn}},\ and\ \bibinfo
  {author} {\bibfnamefont {M.}~\bibnamefont
  {Weidem{\ifmmode\ddot{u}\else\"{u}\fi}ller}},\ }\bibfield  {title} {\bibinfo
  {title} {{Glassy Dynamics in a Disordered Heisenberg Quantum Spin System}},\
  }\href {https://doi.org/10.1103/PhysRevX.11.011011} {\bibfield  {journal}
  {\bibinfo  {journal} {Phys. Rev. X}\ }\textbf {\bibinfo {volume} {11}},\
  \bibinfo {pages} {011011} (\bibinfo {year} {2021})}\BibitemShut {NoStop}%
\bibitem [{\citenamefont {Yan}\ \emph {et~al.}(2013)\citenamefont {Yan},
  \citenamefont {Moses}, \citenamefont {Gadway}, \citenamefont {Covey},
  \citenamefont {Hazzard}, \citenamefont {Rey}, \citenamefont {Jin},\ and\
  \citenamefont {Ye}}]{yan2013observation}%
  \BibitemOpen
  \bibfield  {author} {\bibinfo {author} {\bibfnamefont {B.}~\bibnamefont
  {Yan}}, \bibinfo {author} {\bibfnamefont {S.~A.}\ \bibnamefont {Moses}},
  \bibinfo {author} {\bibfnamefont {B.}~\bibnamefont {Gadway}}, \bibinfo
  {author} {\bibfnamefont {J.~P.}\ \bibnamefont {Covey}}, \bibinfo {author}
  {\bibfnamefont {K.~R.}\ \bibnamefont {Hazzard}}, \bibinfo {author}
  {\bibfnamefont {A.~M.}\ \bibnamefont {Rey}}, \bibinfo {author} {\bibfnamefont
  {D.~S.}\ \bibnamefont {Jin}},\ and\ \bibinfo {author} {\bibfnamefont
  {J.}~\bibnamefont {Ye}},\ }\bibfield  {title} {\bibinfo {title} {Observation
  of dipolar spin-exchange interactions with lattice-confined polar
  molecules},\ }\href {https://doi.org/10.1038/nature12483} {\bibfield
  {journal} {\bibinfo  {journal} {Nature}\ }\textbf {\bibinfo {volume} {501}},\
  \bibinfo {pages} {521} (\bibinfo {year} {2013})}\BibitemShut {NoStop}%
\bibitem [{\citenamefont {Bohn}\ \emph {et~al.}(2017)\citenamefont {Bohn},
  \citenamefont {Rey},\ and\ \citenamefont {Ye}}]{Bohn_Science_357_2017}%
  \BibitemOpen
  \bibfield  {author} {\bibinfo {author} {\bibfnamefont {J.~L.}\ \bibnamefont
  {Bohn}}, \bibinfo {author} {\bibfnamefont {A.~M.}\ \bibnamefont {Rey}},\ and\
  \bibinfo {author} {\bibfnamefont {J.}~\bibnamefont {Ye}},\ }\bibfield
  {title} {\bibinfo {title} {Cold molecules: Progress in quantum engineering of
  chemistry and quantum matter},\ }\href
  {https://doi.org/10.1126/science.aam6299} {\bibfield  {journal} {\bibinfo
  {journal} {Science}\ }\textbf {\bibinfo {volume} {357}},\ \bibinfo {pages}
  {1002} (\bibinfo {year} {2017})}\BibitemShut {NoStop}%
\bibitem [{\citenamefont {Moses}\ \emph {et~al.}(2017)\citenamefont {Moses},
  \citenamefont {Covey}, \citenamefont {Miecnikowski}, \citenamefont {Jin},\
  and\ \citenamefont {Ye}}]{moses2017new}%
  \BibitemOpen
  \bibfield  {author} {\bibinfo {author} {\bibfnamefont {S.~A.}\ \bibnamefont
  {Moses}}, \bibinfo {author} {\bibfnamefont {J.~P.}\ \bibnamefont {Covey}},
  \bibinfo {author} {\bibfnamefont {M.~T.}\ \bibnamefont {Miecnikowski}},
  \bibinfo {author} {\bibfnamefont {D.~S.}\ \bibnamefont {Jin}},\ and\ \bibinfo
  {author} {\bibfnamefont {J.}~\bibnamefont {Ye}},\ }\bibfield  {title}
  {\bibinfo {title} {{New frontiers for quantum gases of polar molecules}},\
  }\href {https://doi.org/10.1038/nphys3985} {\bibfield  {journal} {\bibinfo
  {journal} {Nat. Phys.}\ }\textbf {\bibinfo {volume} {13}},\ \bibinfo {pages}
  {13} (\bibinfo {year} {2017})}\BibitemShut {NoStop}%
\bibitem [{\citenamefont {Kaufman}\ and\ \citenamefont
  {Ni}(2021)}]{NatPhysKaufman2021}%
  \BibitemOpen
  \bibfield  {author} {\bibinfo {author} {\bibfnamefont {A.~M.}\ \bibnamefont
  {Kaufman}}\ and\ \bibinfo {author} {\bibfnamefont {K.-K.}\ \bibnamefont
  {Ni}},\ }\bibfield  {title} {\bibinfo {title} {Quantum science with optical
  tweezer arrays of ultracold atoms and molecules},\ }\href
  {https://doi.org/10.1038/s41567-021-01357-2} {\bibfield  {journal} {\bibinfo
  {journal} {Nature Physics}\ }\textbf {\bibinfo {volume} {17}},\ \bibinfo
  {pages} {1324} (\bibinfo {year} {2021})}\BibitemShut {NoStop}%
\bibitem [{\citenamefont {Bao}\ \emph {et~al.}(2022)\citenamefont {Bao},
  \citenamefont {Yu}, \citenamefont {Anderegg}, \citenamefont {Chae},
  \citenamefont {Ketterle}, \citenamefont {Ni},\ and\ \citenamefont
  {Doyle}}]{bao2022dipolar}%
  \BibitemOpen
  \bibfield  {author} {\bibinfo {author} {\bibfnamefont {Y.}~\bibnamefont
  {Bao}}, \bibinfo {author} {\bibfnamefont {S.~S.}\ \bibnamefont {Yu}},
  \bibinfo {author} {\bibfnamefont {L.}~\bibnamefont {Anderegg}}, \bibinfo
  {author} {\bibfnamefont {E.}~\bibnamefont {Chae}}, \bibinfo {author}
  {\bibfnamefont {W.}~\bibnamefont {Ketterle}}, \bibinfo {author}
  {\bibfnamefont {K.-K.}\ \bibnamefont {Ni}},\ and\ \bibinfo {author}
  {\bibfnamefont {J.~M.}\ \bibnamefont {Doyle}},\ }\href@noop {} {\bibinfo
  {title} {Dipolar spin-exchange and entanglement between molecules in an
  optical tweezer array}} (\bibinfo {year} {2022}),\ \Eprint
  {https://arxiv.org/abs/2211.09780} {arXiv:2211.09780 [physics.atom-ph]}
  \BibitemShut {NoStop}%
\bibitem [{\citenamefont {Holland}\ \emph {et~al.}(2022)\citenamefont
  {Holland}, \citenamefont {Lu},\ and\ \citenamefont
  {Cheuk}}]{holland2022ondemand}%
  \BibitemOpen
  \bibfield  {author} {\bibinfo {author} {\bibfnamefont {C.~M.}\ \bibnamefont
  {Holland}}, \bibinfo {author} {\bibfnamefont {Y.}~\bibnamefont {Lu}},\ and\
  \bibinfo {author} {\bibfnamefont {L.~W.}\ \bibnamefont {Cheuk}},\ }\href@noop
  {} {\bibinfo {title} {On-demand entanglement of molecules in a reconfigurable
  optical tweezer array}} (\bibinfo {year} {2022}),\ \Eprint
  {https://arxiv.org/abs/2210.06309} {arXiv:2210.06309 [cond-mat.quant-gas]}
  \BibitemShut {NoStop}%
\bibitem [{\citenamefont {Zhang}\ \emph {et~al.}(2022)\citenamefont {Zhang},
  \citenamefont {Picard}, \citenamefont {Cairncross}, \citenamefont {Wang},
  \citenamefont {Yu}, \citenamefont {Fang},\ and\ \citenamefont
  {Ni}}]{zhang2022optical}%
  \BibitemOpen
  \bibfield  {author} {\bibinfo {author} {\bibfnamefont {J.~T.}\ \bibnamefont
  {Zhang}}, \bibinfo {author} {\bibfnamefont {L.~R.~B.}\ \bibnamefont
  {Picard}}, \bibinfo {author} {\bibfnamefont {W.~B.}\ \bibnamefont
  {Cairncross}}, \bibinfo {author} {\bibfnamefont {K.}~\bibnamefont {Wang}},
  \bibinfo {author} {\bibfnamefont {Y.}~\bibnamefont {Yu}}, \bibinfo {author}
  {\bibfnamefont {F.}~\bibnamefont {Fang}},\ and\ \bibinfo {author}
  {\bibfnamefont {K.-K.}\ \bibnamefont {Ni}},\ }\bibfield  {title} {\bibinfo
  {title} {{An optical tweezer array of ground-state polar molecules}},\ }\href
  {https://doi.org/10.1088/2058-9565/ac676c} {\bibfield  {journal} {\bibinfo
  {journal} {Quantum Sci. Technol.}\ }\textbf {\bibinfo {volume} {7}},\
  \bibinfo {pages} {035006} (\bibinfo {year} {2022})}\BibitemShut {NoStop}%
\bibitem [{\citenamefont {Christakis}\ \emph
  {et~al.}(2023{\natexlab{a}})\citenamefont {Christakis}, \citenamefont
  {Rosenberg}, \citenamefont {Raj}, \citenamefont {Chi}, \citenamefont
  {Morningstar}, \citenamefont {Huse}, \citenamefont {Yan},\ and\ \citenamefont
  {Bakr}}]{christakis2022probing}%
  \BibitemOpen
  \bibfield  {author} {\bibinfo {author} {\bibfnamefont {L.}~\bibnamefont
  {Christakis}}, \bibinfo {author} {\bibfnamefont {J.~S.}\ \bibnamefont
  {Rosenberg}}, \bibinfo {author} {\bibfnamefont {R.}~\bibnamefont {Raj}},
  \bibinfo {author} {\bibfnamefont {S.}~\bibnamefont {Chi}}, \bibinfo {author}
  {\bibfnamefont {A.}~\bibnamefont {Morningstar}}, \bibinfo {author}
  {\bibfnamefont {D.~A.}\ \bibnamefont {Huse}}, \bibinfo {author}
  {\bibfnamefont {Z.~Z.}\ \bibnamefont {Yan}},\ and\ \bibinfo {author}
  {\bibfnamefont {W.~S.}\ \bibnamefont {Bakr}},\ }\bibfield  {title} {\bibinfo
  {title} {Probing site-resolved correlations in a spin system of ultracold
  molecules},\ }\href {https://doi.org/10.1038/s41586-022-05558-4} {\bibfield
  {journal} {\bibinfo  {journal} {Nature}\ }\textbf {\bibinfo {volume} {614}},\
  \bibinfo {pages} {64} (\bibinfo {year} {2023}{\natexlab{a}})}\BibitemShut
  {NoStop}%
\bibitem [{\citenamefont {Atala}\ \emph {et~al.}(2014)\citenamefont {Atala},
  \citenamefont {Aidelsburger}, \citenamefont {Lohse}, \citenamefont
  {Barreiro}, \citenamefont {Paredes},\ and\ \citenamefont
  {Bloch}}]{Atala2014}%
  \BibitemOpen
  \bibfield  {author} {\bibinfo {author} {\bibfnamefont {M.}~\bibnamefont
  {Atala}}, \bibinfo {author} {\bibfnamefont {M.}~\bibnamefont {Aidelsburger}},
  \bibinfo {author} {\bibfnamefont {M.}~\bibnamefont {Lohse}}, \bibinfo
  {author} {\bibfnamefont {J.~T.}\ \bibnamefont {Barreiro}}, \bibinfo {author}
  {\bibfnamefont {B.}~\bibnamefont {Paredes}},\ and\ \bibinfo {author}
  {\bibfnamefont {I.}~\bibnamefont {Bloch}},\ }\bibfield  {title} {\bibinfo
  {title} {Observation of chiral currents with ultracold atoms in bosonic
  ladders},\ }\href {https://doi.org/10.1038/nphys2998} {\bibfield  {journal}
  {\bibinfo  {journal} {Nature Physics}\ }\textbf {\bibinfo {volume} {10}},\
  \bibinfo {pages} {588} (\bibinfo {year} {2014})}\BibitemShut {NoStop}%
\bibitem [{\citenamefont {Ye}\ \emph {et~al.}(2019)\citenamefont {Ye},
  \citenamefont {Ge}, \citenamefont {Wu}, \citenamefont {Wang}, \citenamefont
  {Gong}, \citenamefont {Zhang}, \citenamefont {Zhu}, \citenamefont {Yang},
  \citenamefont {Li}, \citenamefont {Liang}, \citenamefont {Lin}, \citenamefont
  {Xu}, \citenamefont {Guo}, \citenamefont {Sun}, \citenamefont {Cheng},
  \citenamefont {Ma}, \citenamefont {Meng}, \citenamefont {Deng}, \citenamefont
  {Rong}, \citenamefont {Lu}, \citenamefont {Peng}, \citenamefont {Fan},
  \citenamefont {Zhu},\ and\ \citenamefont {Pan}}]{Ye2019}%
  \BibitemOpen
  \bibfield  {author} {\bibinfo {author} {\bibfnamefont {Y.}~\bibnamefont
  {Ye}}, \bibinfo {author} {\bibfnamefont {Z.-Y.}\ \bibnamefont {Ge}}, \bibinfo
  {author} {\bibfnamefont {Y.}~\bibnamefont {Wu}}, \bibinfo {author}
  {\bibfnamefont {S.}~\bibnamefont {Wang}}, \bibinfo {author} {\bibfnamefont
  {M.}~\bibnamefont {Gong}}, \bibinfo {author} {\bibfnamefont {Y.-R.}\
  \bibnamefont {Zhang}}, \bibinfo {author} {\bibfnamefont {Q.}~\bibnamefont
  {Zhu}}, \bibinfo {author} {\bibfnamefont {R.}~\bibnamefont {Yang}}, \bibinfo
  {author} {\bibfnamefont {S.}~\bibnamefont {Li}}, \bibinfo {author}
  {\bibfnamefont {F.}~\bibnamefont {Liang}}, \bibinfo {author} {\bibfnamefont
  {J.}~\bibnamefont {Lin}}, \bibinfo {author} {\bibfnamefont {Y.}~\bibnamefont
  {Xu}}, \bibinfo {author} {\bibfnamefont {C.}~\bibnamefont {Guo}}, \bibinfo
  {author} {\bibfnamefont {L.}~\bibnamefont {Sun}}, \bibinfo {author}
  {\bibfnamefont {C.}~\bibnamefont {Cheng}}, \bibinfo {author} {\bibfnamefont
  {N.}~\bibnamefont {Ma}}, \bibinfo {author} {\bibfnamefont {Z.~Y.}\
  \bibnamefont {Meng}}, \bibinfo {author} {\bibfnamefont {H.}~\bibnamefont
  {Deng}}, \bibinfo {author} {\bibfnamefont {H.}~\bibnamefont {Rong}}, \bibinfo
  {author} {\bibfnamefont {C.-Y.}\ \bibnamefont {Lu}}, \bibinfo {author}
  {\bibfnamefont {C.-Z.}\ \bibnamefont {Peng}}, \bibinfo {author}
  {\bibfnamefont {H.}~\bibnamefont {Fan}}, \bibinfo {author} {\bibfnamefont
  {X.}~\bibnamefont {Zhu}},\ and\ \bibinfo {author} {\bibfnamefont {J.-W.}\
  \bibnamefont {Pan}},\ }\bibfield  {title} {\bibinfo {title} {{Propagation and
  Localization of Collective Excitations on a 24-Qubit Superconducting
  Processor}},\ }\href {https://doi.org/10.1103/PhysRevLett.123.050502}
  {\bibfield  {journal} {\bibinfo  {journal} {Phys. Rev. Lett.}\ }\textbf
  {\bibinfo {volume} {123}},\ \bibinfo {pages} {050502} (\bibinfo {year}
  {2019})}\BibitemShut {NoStop}%
\bibitem [{\citenamefont {Gall}\ \emph {et~al.}(2021)\citenamefont {Gall},
  \citenamefont {Wurz}, \citenamefont {Samland}, \citenamefont {Chan},\ and\
  \citenamefont {Köhl}}]{Gall2021}%
  \BibitemOpen
  \bibfield  {author} {\bibinfo {author} {\bibfnamefont {M.}~\bibnamefont
  {Gall}}, \bibinfo {author} {\bibfnamefont {N.}~\bibnamefont {Wurz}}, \bibinfo
  {author} {\bibfnamefont {J.}~\bibnamefont {Samland}}, \bibinfo {author}
  {\bibfnamefont {C.~F.}\ \bibnamefont {Chan}},\ and\ \bibinfo {author}
  {\bibfnamefont {M.}~\bibnamefont {Köhl}},\ }\bibfield  {title} {\bibinfo
  {title} {Competing magnetic orders in a bilayer hubbard model with ultracold
  atoms},\ }\href {https://doi.org/10.1038/s41586-020-03058-x} {\bibfield
  {journal} {\bibinfo  {journal} {Nature}\ }\textbf {\bibinfo {volume} {589}},\
  \bibinfo {pages} {40} (\bibinfo {year} {2021})}\BibitemShut {NoStop}%
\bibitem [{\citenamefont {Hirthe}\ \emph
  {et~al.}(2023{\natexlab{b}})\citenamefont {Hirthe}, \citenamefont {Chalopin},
  \citenamefont {Bourgund}, \citenamefont {Bojović}, \citenamefont {Bohrdt},
  \citenamefont {Demler}, \citenamefont {Grusdt}, \citenamefont {Bloch},\ and\
  \citenamefont {Hilker}}]{Hirthe2023}%
  \BibitemOpen
  \bibfield  {author} {\bibinfo {author} {\bibfnamefont {S.}~\bibnamefont
  {Hirthe}}, \bibinfo {author} {\bibfnamefont {T.}~\bibnamefont {Chalopin}},
  \bibinfo {author} {\bibfnamefont {D.}~\bibnamefont {Bourgund}}, \bibinfo
  {author} {\bibfnamefont {P.}~\bibnamefont {Bojović}}, \bibinfo {author}
  {\bibfnamefont {A.}~\bibnamefont {Bohrdt}}, \bibinfo {author} {\bibfnamefont
  {E.}~\bibnamefont {Demler}}, \bibinfo {author} {\bibfnamefont
  {F.}~\bibnamefont {Grusdt}}, \bibinfo {author} {\bibfnamefont
  {I.}~\bibnamefont {Bloch}},\ and\ \bibinfo {author} {\bibfnamefont {T.~A.}\
  \bibnamefont {Hilker}},\ }\bibfield  {title} {\bibinfo {title} {Magnetically
  mediated hole pairing in fermionic ladders of ultracold atoms},\ }\href
  {https://doi.org/10.1038/s41586-022-05437-y} {\bibfield  {journal} {\bibinfo
  {journal} {Nature}\ }\textbf {\bibinfo {volume} {613}},\ \bibinfo {pages}
  {463} (\bibinfo {year} {2023}{\natexlab{b}})}\BibitemShut {NoStop}%
\bibitem [{\citenamefont {Wienand}\ \emph {et~al.}(2023)\citenamefont
  {Wienand}, \citenamefont {Karch}, \citenamefont {Impertro}, \citenamefont
  {Schweizer}, \citenamefont {McCulloch}, \citenamefont {Vasseur},
  \citenamefont {Gopalakrishnan}, \citenamefont {Aidelsburger},\ and\
  \citenamefont {Bloch}}]{wienand2023emergence}%
  \BibitemOpen
  \bibfield  {author} {\bibinfo {author} {\bibfnamefont {J.~F.}\ \bibnamefont
  {Wienand}}, \bibinfo {author} {\bibfnamefont {S.}~\bibnamefont {Karch}},
  \bibinfo {author} {\bibfnamefont {A.}~\bibnamefont {Impertro}}, \bibinfo
  {author} {\bibfnamefont {C.}~\bibnamefont {Schweizer}}, \bibinfo {author}
  {\bibfnamefont {E.}~\bibnamefont {McCulloch}}, \bibinfo {author}
  {\bibfnamefont {R.}~\bibnamefont {Vasseur}}, \bibinfo {author} {\bibfnamefont
  {S.}~\bibnamefont {Gopalakrishnan}}, \bibinfo {author} {\bibfnamefont
  {M.}~\bibnamefont {Aidelsburger}},\ and\ \bibinfo {author} {\bibfnamefont
  {I.}~\bibnamefont {Bloch}},\ }\href@noop {} {\bibinfo {title} {Emergence of
  fluctuating hydrodynamics in chaotic quantum systems}} (\bibinfo {year}
  {2023}),\ \Eprint {https://arxiv.org/abs/2306.11457} {arXiv:2306.11457
  [cond-mat.quant-gas]} \BibitemShut {NoStop}%
\bibitem [{\citenamefont {Tobias}\ \emph {et~al.}(2022)\citenamefont {Tobias},
  \citenamefont {Matsuda}, \citenamefont {Li}, \citenamefont {Miller},
  \citenamefont {Carroll}, \citenamefont {Bilitewski}, \citenamefont {Rey},\
  and\ \citenamefont {Ye}}]{Tobias_Science_375_2022}%
  \BibitemOpen
  \bibfield  {author} {\bibinfo {author} {\bibfnamefont {W.~G.}\ \bibnamefont
  {Tobias}}, \bibinfo {author} {\bibfnamefont {K.}~\bibnamefont {Matsuda}},
  \bibinfo {author} {\bibfnamefont {J.-R.}\ \bibnamefont {Li}}, \bibinfo
  {author} {\bibfnamefont {C.}~\bibnamefont {Miller}}, \bibinfo {author}
  {\bibfnamefont {A.~N.}\ \bibnamefont {Carroll}}, \bibinfo {author}
  {\bibfnamefont {T.}~\bibnamefont {Bilitewski}}, \bibinfo {author}
  {\bibfnamefont {A.~M.}\ \bibnamefont {Rey}},\ and\ \bibinfo {author}
  {\bibfnamefont {J.}~\bibnamefont {Ye}},\ }\bibfield  {title} {\bibinfo
  {title} {Reactions between layer-resolved molecules mediated by dipolar spin
  exchange},\ }\href {https://doi.org/10.1126/science.abn8525} {\bibfield
  {journal} {\bibinfo  {journal} {Science}\ }\textbf {\bibinfo {volume}
  {375}},\ \bibinfo {pages} {1299} (\bibinfo {year} {2022})}\BibitemShut
  {NoStop}%
\bibitem [{\citenamefont {Bilitewski}\ and\ \citenamefont
  {Rey}(2023)}]{bilitewski2023manipulating}%
  \BibitemOpen
  \bibfield  {author} {\bibinfo {author} {\bibfnamefont {T.}~\bibnamefont
  {Bilitewski}}\ and\ \bibinfo {author} {\bibfnamefont {A.~M.}\ \bibnamefont
  {Rey}},\ }\bibfield  {title} {\bibinfo {title} {Manipulating growth and
  propagation of correlations in dipolar multilayers: From pair production to
  bosonic kitaev models},\ }\href
  {https://doi.org/10.1103/PhysRevLett.131.053001} {\bibfield  {journal}
  {\bibinfo  {journal} {Phys. Rev. Lett.}\ }\textbf {\bibinfo {volume} {131}},\
  \bibinfo {pages} {053001} (\bibinfo {year} {2023})}\BibitemShut {NoStop}%
\bibitem [{\citenamefont {Bilitewski}\ \emph {et~al.}(2023)\citenamefont
  {Bilitewski}, \citenamefont {Dom\'{\i}nguez-Castro}, \citenamefont
  {Wellnitz}, \citenamefont {Rey},\ and\ \citenamefont
  {Santos}}]{bilitewski2023momentumselective}%
  \BibitemOpen
  \bibfield  {author} {\bibinfo {author} {\bibfnamefont {T.}~\bibnamefont
  {Bilitewski}}, \bibinfo {author} {\bibfnamefont {G.~A.}\ \bibnamefont
  {Dom\'{\i}nguez-Castro}}, \bibinfo {author} {\bibfnamefont {D.}~\bibnamefont
  {Wellnitz}}, \bibinfo {author} {\bibfnamefont {A.~M.}\ \bibnamefont {Rey}},\
  and\ \bibinfo {author} {\bibfnamefont {L.}~\bibnamefont {Santos}},\
  }\bibfield  {title} {\bibinfo {title} {Tunable momentum pair creation of spin
  excitations in dipolar bilayers},\ }\href
  {https://doi.org/10.1103/PhysRevA.108.013313} {\bibfield  {journal} {\bibinfo
   {journal} {Phys. Rev. A}\ }\textbf {\bibinfo {volume} {108}},\ \bibinfo
  {pages} {013313} (\bibinfo {year} {2023})}\BibitemShut {NoStop}%
\bibitem [{\citenamefont {Agarwal}(2013)}]{agarwal2013quantum}%
  \BibitemOpen
  \bibfield  {author} {\bibinfo {author} {\bibfnamefont {G.}~\bibnamefont
  {Agarwal}},\ }\href {https://books.google.com/books?id=7KKw\_XIYaioC} {\emph
  {\bibinfo {title} {Quantum Optics}}},\ Quantum Optics\ (\bibinfo  {publisher}
  {Cambridge University Press},\ \bibinfo {year} {2013})\BibitemShut {NoStop}%
\bibitem [{\citenamefont {Tian}\ \emph {et~al.}(2018)\citenamefont {Tian},
  \citenamefont {Ch\"a},\ and\ \citenamefont {Fischer}}]{PhysRevA.97.063611}%
  \BibitemOpen
  \bibfield  {author} {\bibinfo {author} {\bibfnamefont {Z.}~\bibnamefont
  {Tian}}, \bibinfo {author} {\bibfnamefont {S.-Y.}\ \bibnamefont {Ch\"a}},\
  and\ \bibinfo {author} {\bibfnamefont {U.~R.}\ \bibnamefont {Fischer}},\
  }\bibfield  {title} {\bibinfo {title} {Roton entanglement in quenched dipolar
  bose-einstein condensates},\ }\href
  {https://doi.org/10.1103/PhysRevA.97.063611} {\bibfield  {journal} {\bibinfo
  {journal} {Phys. Rev. A}\ }\textbf {\bibinfo {volume} {97}},\ \bibinfo
  {pages} {063611} (\bibinfo {year} {2018})}\BibitemShut {NoStop}%
\bibitem [{\citenamefont {Rigol}\ \emph {et~al.}(2007)\citenamefont {Rigol},
  \citenamefont {Dunjko}, \citenamefont {Yurovsky},\ and\ \citenamefont
  {Olshanii}}]{Rigol2007}%
  \BibitemOpen
  \bibfield  {author} {\bibinfo {author} {\bibfnamefont {M.}~\bibnamefont
  {Rigol}}, \bibinfo {author} {\bibfnamefont {V.}~\bibnamefont {Dunjko}},
  \bibinfo {author} {\bibfnamefont {V.}~\bibnamefont {Yurovsky}},\ and\
  \bibinfo {author} {\bibfnamefont {M.}~\bibnamefont {Olshanii}},\ }\bibfield
  {title} {\bibinfo {title} {Relaxation in a completely integrable many-body
  quantum system: An ab initio study of the dynamics of the highly excited
  states of 1d lattice hard-core bosons},\ }\href
  {https://doi.org/10.1103/PhysRevLett.98.050405} {\bibfield  {journal}
  {\bibinfo  {journal} {Phys. Rev. Lett.}\ }\textbf {\bibinfo {volume} {98}},\
  \bibinfo {pages} {050405} (\bibinfo {year} {2007})}\BibitemShut {NoStop}%
\bibitem [{\citenamefont {Abanin}\ \emph {et~al.}(2019)\citenamefont {Abanin},
  \citenamefont {Altman}, \citenamefont {Bloch},\ and\ \citenamefont
  {Serbyn}}]{Abanin2019}%
  \BibitemOpen
  \bibfield  {author} {\bibinfo {author} {\bibfnamefont {D.~A.}\ \bibnamefont
  {Abanin}}, \bibinfo {author} {\bibfnamefont {E.}~\bibnamefont {Altman}},
  \bibinfo {author} {\bibfnamefont {I.}~\bibnamefont {Bloch}},\ and\ \bibinfo
  {author} {\bibfnamefont {M.}~\bibnamefont {Serbyn}},\ }\bibfield  {title}
  {\bibinfo {title} {Colloquium: Many-body localization, thermalization, and
  entanglement},\ }\href {https://doi.org/10.1103/RevModPhys.91.021001}
  {\bibfield  {journal} {\bibinfo  {journal} {Rev. Mod. Phys.}\ }\textbf
  {\bibinfo {volume} {91}},\ \bibinfo {pages} {021001} (\bibinfo {year}
  {2019})}\BibitemShut {NoStop}%
\bibitem [{\citenamefont {Luca~D'Alessio}\ and\ \citenamefont
  {Rigol}(2016)}]{Dalessio2016}%
  \BibitemOpen
  \bibfield  {author} {\bibinfo {author} {\bibfnamefont {A.~P.}\ \bibnamefont
  {Luca~D'Alessio}, \bibfnamefont {Yariv~Kafri}}\ and\ \bibinfo {author}
  {\bibfnamefont {M.}~\bibnamefont {Rigol}},\ }\bibfield  {title} {\bibinfo
  {title} {From quantum chaos and eigenstate thermalization to statistical
  mechanics and thermodynamics},\ }\href
  {https://doi.org/10.1080/00018732.2016.1198134} {\bibfield  {journal}
  {\bibinfo  {journal} {Advances in Physics}\ }\textbf {\bibinfo {volume}
  {65}},\ \bibinfo {pages} {239} (\bibinfo {year} {2016})}\BibitemShut
  {NoStop}%
\bibitem [{\citenamefont {Mallayya}\ \emph {et~al.}(2019)\citenamefont
  {Mallayya}, \citenamefont {Rigol},\ and\ \citenamefont
  {De~Roeck}}]{Mallayya2019}%
  \BibitemOpen
  \bibfield  {author} {\bibinfo {author} {\bibfnamefont {K.}~\bibnamefont
  {Mallayya}}, \bibinfo {author} {\bibfnamefont {M.}~\bibnamefont {Rigol}},\
  and\ \bibinfo {author} {\bibfnamefont {W.}~\bibnamefont {De~Roeck}},\
  }\bibfield  {title} {\bibinfo {title} {Prethermalization and thermalization
  in isolated quantum systems},\ }\href
  {https://doi.org/10.1103/PhysRevX.9.021027} {\bibfield  {journal} {\bibinfo
  {journal} {Phys. Rev. X}\ }\textbf {\bibinfo {volume} {9}},\ \bibinfo {pages}
  {021027} (\bibinfo {year} {2019})}\BibitemShut {NoStop}%
\bibitem [{\citenamefont {Bastianello}\ \emph {et~al.}(2021)\citenamefont
  {Bastianello}, \citenamefont {Luca},\ and\ \citenamefont
  {Vasseur}}]{Bastianello2021}%
  \BibitemOpen
  \bibfield  {author} {\bibinfo {author} {\bibfnamefont {A.}~\bibnamefont
  {Bastianello}}, \bibinfo {author} {\bibfnamefont {A.~D.}\ \bibnamefont
  {Luca}},\ and\ \bibinfo {author} {\bibfnamefont {R.}~\bibnamefont
  {Vasseur}},\ }\bibfield  {title} {\bibinfo {title} {Hydrodynamics of weak
  integrability breaking},\ }\href {https://doi.org/10.1088/1742-5468/ac26b2}
  {\bibfield  {journal} {\bibinfo  {journal} {Journal of Statistical Mechanics:
  Theory and Experiment}\ }\textbf {\bibinfo {volume} {2021}},\ \bibinfo
  {pages} {114003} (\bibinfo {year} {2021})}\BibitemShut {NoStop}%
\bibitem [{\citenamefont {Coleman}(1977)}]{coleman1977}%
  \BibitemOpen
  \bibfield  {author} {\bibinfo {author} {\bibfnamefont {S.}~\bibnamefont
  {Coleman}},\ }\bibfield  {title} {\bibinfo {title} {Fate of the false vacuum:
  Semiclassical theory},\ }\href {https://doi.org/10.1103/PhysRevD.15.2929}
  {\bibfield  {journal} {\bibinfo  {journal} {Phys. Rev. D}\ }\textbf {\bibinfo
  {volume} {15}},\ \bibinfo {pages} {2929} (\bibinfo {year}
  {1977})}\BibitemShut {NoStop}%
\bibitem [{\citenamefont {Lagnese}\ \emph {et~al.}(2021)\citenamefont
  {Lagnese}, \citenamefont {Surace}, \citenamefont {Kormos},\ and\
  \citenamefont {Calabrese}}]{Lagnese2021}%
  \BibitemOpen
  \bibfield  {author} {\bibinfo {author} {\bibfnamefont {G.}~\bibnamefont
  {Lagnese}}, \bibinfo {author} {\bibfnamefont {F.~M.}\ \bibnamefont {Surace}},
  \bibinfo {author} {\bibfnamefont {M.}~\bibnamefont {Kormos}},\ and\ \bibinfo
  {author} {\bibfnamefont {P.}~\bibnamefont {Calabrese}},\ }\bibfield  {title}
  {\bibinfo {title} {False vacuum decay in quantum spin chains},\ }\href
  {https://doi.org/10.1103/PhysRevB.104.L201106} {\bibfield  {journal}
  {\bibinfo  {journal} {Phys. Rev. B}\ }\textbf {\bibinfo {volume} {104}},\
  \bibinfo {pages} {L201106} (\bibinfo {year} {2021})}\BibitemShut {NoStop}%
\bibitem [{\citenamefont {Milsted}\ \emph {et~al.}(2022)\citenamefont
  {Milsted}, \citenamefont {Liu}, \citenamefont {Preskill},\ and\ \citenamefont
  {Vidal}}]{PRXQuantum.3.020316}%
  \BibitemOpen
  \bibfield  {author} {\bibinfo {author} {\bibfnamefont {A.}~\bibnamefont
  {Milsted}}, \bibinfo {author} {\bibfnamefont {J.}~\bibnamefont {Liu}},
  \bibinfo {author} {\bibfnamefont {J.}~\bibnamefont {Preskill}},\ and\
  \bibinfo {author} {\bibfnamefont {G.}~\bibnamefont {Vidal}},\ }\bibfield
  {title} {\bibinfo {title} {Collisions of false-vacuum bubble walls in a
  quantum spin chain},\ }\href {https://doi.org/10.1103/PRXQuantum.3.020316}
  {\bibfield  {journal} {\bibinfo  {journal} {PRX Quantum}\ }\textbf {\bibinfo
  {volume} {3}},\ \bibinfo {pages} {020316} (\bibinfo {year}
  {2022})}\BibitemShut {NoStop}%
\bibitem [{\citenamefont {Zenesini}\ \emph {et~al.}(2024)\citenamefont
  {Zenesini}, \citenamefont {Berti}, \citenamefont {Cominotti}, \citenamefont
  {Rogora}, \citenamefont {Moss}, \citenamefont {Billam}, \citenamefont
  {Carusotto}, \citenamefont {Lamporesi}, \citenamefont {Recati},\ and\
  \citenamefont {Ferrari}}]{Zenesini2024}%
  \BibitemOpen
  \bibfield  {author} {\bibinfo {author} {\bibfnamefont {A.}~\bibnamefont
  {Zenesini}}, \bibinfo {author} {\bibfnamefont {A.}~\bibnamefont {Berti}},
  \bibinfo {author} {\bibfnamefont {R.}~\bibnamefont {Cominotti}}, \bibinfo
  {author} {\bibfnamefont {C.}~\bibnamefont {Rogora}}, \bibinfo {author}
  {\bibfnamefont {I.~G.}\ \bibnamefont {Moss}}, \bibinfo {author}
  {\bibfnamefont {T.~P.}\ \bibnamefont {Billam}}, \bibinfo {author}
  {\bibfnamefont {I.}~\bibnamefont {Carusotto}}, \bibinfo {author}
  {\bibfnamefont {G.}~\bibnamefont {Lamporesi}}, \bibinfo {author}
  {\bibfnamefont {A.}~\bibnamefont {Recati}},\ and\ \bibinfo {author}
  {\bibfnamefont {G.}~\bibnamefont {Ferrari}},\ }\bibfield  {title} {\bibinfo
  {title} {False vacuum decay via bubble formation in ferromagnetic
  superfluids},\ }\href {https://doi.org/10.1038/s41567-023-02345-4} {\bibfield
   {journal} {\bibinfo  {journal} {Nature Physics}\ }\textbf {\bibinfo {volume}
  {20}},\ \bibinfo {pages} {558–563} (\bibinfo {year} {2024})}\BibitemShut
  {NoStop}%
\bibitem [{\citenamefont {Steinigeweg}\ \emph {et~al.}(2014)\citenamefont
  {Steinigeweg}, \citenamefont {Heidrich-Meisner}, \citenamefont {Gemmer},
  \citenamefont {Michielsen},\ and\ \citenamefont {De~Raedt}}]{Steinigweg2014}%
  \BibitemOpen
  \bibfield  {author} {\bibinfo {author} {\bibfnamefont {R.}~\bibnamefont
  {Steinigeweg}}, \bibinfo {author} {\bibfnamefont {F.}~\bibnamefont
  {Heidrich-Meisner}}, \bibinfo {author} {\bibfnamefont {J.}~\bibnamefont
  {Gemmer}}, \bibinfo {author} {\bibfnamefont {K.}~\bibnamefont {Michielsen}},\
  and\ \bibinfo {author} {\bibfnamefont {H.}~\bibnamefont {De~Raedt}},\
  }\bibfield  {title} {\bibinfo {title} {Scaling of diffusion constants in the
  spin-$\frac{1}{2}$ xx ladder},\ }\href
  {https://doi.org/10.1103/PhysRevB.90.094417} {\bibfield  {journal} {\bibinfo
  {journal} {Phys. Rev. B}\ }\textbf {\bibinfo {volume} {90}},\ \bibinfo
  {pages} {094417} (\bibinfo {year} {2014})}\BibitemShut {NoStop}%
\bibitem [{\citenamefont {Rakovszky}\ \emph {et~al.}(2022)\citenamefont
  {Rakovszky}, \citenamefont {von Keyserlingk},\ and\ \citenamefont
  {Pollmann}}]{Rakovszky2022}%
  \BibitemOpen
  \bibfield  {author} {\bibinfo {author} {\bibfnamefont {T.}~\bibnamefont
  {Rakovszky}}, \bibinfo {author} {\bibfnamefont {C.~W.}\ \bibnamefont {von
  Keyserlingk}},\ and\ \bibinfo {author} {\bibfnamefont {F.}~\bibnamefont
  {Pollmann}},\ }\bibfield  {title} {\bibinfo {title} {Dissipation-assisted
  operator evolution method for capturing hydrodynamic transport},\ }\href
  {https://doi.org/10.1103/PhysRevB.105.075131} {\bibfield  {journal} {\bibinfo
   {journal} {Phys. Rev. B}\ }\textbf {\bibinfo {volume} {105}},\ \bibinfo
  {pages} {075131} (\bibinfo {year} {2022})}\BibitemShut {NoStop}%
\bibitem [{SM()}]{SM}%
  \BibitemOpen
  \href@noop {} {}\bibinfo {note} {See the Supplemental Material (at the url
  provided by the publisher) for a detailed discussion of the Bogoliubov
  analysis, the Chebyshev method, the equilibrium distribution, the
  entanglement entropy, and extended numerical results.}\BibitemShut {Stop}%
\bibitem [{Qua()}]{QuantumFluctuations}%
  \BibitemOpen
  \href@noop {} {}\bibinfo {note} {Note that quantum fluctuations are crucial
  to trigger the spin dynamics. In classical spins, the vacuum state would not
  evolve.}\BibitemShut {Stop}%
\bibitem [{\citenamefont {Fehske}\ \emph {et~al.}(2007)\citenamefont {Fehske},
  \citenamefont {Schneider},\ and\ \citenamefont
  {Weisse}}]{fehske2007computational}%
  \BibitemOpen
  \bibfield  {author} {\bibinfo {author} {\bibfnamefont {H.}~\bibnamefont
  {Fehske}}, \bibinfo {author} {\bibfnamefont {R.}~\bibnamefont {Schneider}},\
  and\ \bibinfo {author} {\bibfnamefont {A.}~\bibnamefont {Weisse}},\ }\href
  {https://link.springer.com/book/10.1007/978-3-540-74686-7} {\emph {\bibinfo
  {title} {Computational many-particle physics}}},\ Vol.\ \bibinfo {volume}
  {739}\ (\bibinfo  {publisher} {Springer},\ \bibinfo {year}
  {2007})\BibitemShut {NoStop}%
\bibitem [{DTW()}]{DTWA}%
  \BibitemOpen
  \href@noop {} {}\bibinfo {note} {Although a semiclassical method such as the
  discrete truncated Wigner approach satisfactorily reproduces the short-term
  dynamics, it tends to underestimate the relaxation time, pointing to the
  building up of higher-order correlations.}\BibitemShut {Stop}%
\bibitem [{foo()}]{footnote-angle-window}%
  \BibitemOpen
  \href@noop {} {}\bibinfo {note} {In order to avoid very long time scales
  resulting from the vanishing angular dependence of $\bar{V}$, we restrict the
  long-time analysis to orientations with $0<\theta<0.9\pi$.}\BibitemShut
  {Stop}%
\bibitem [{\citenamefont {Takahashi}\ and\ \citenamefont
  {Umezawa}(1996)}]{TAKAHASHI1996}%
  \BibitemOpen
  \bibfield  {author} {\bibinfo {author} {\bibfnamefont {Y.}~\bibnamefont
  {Takahashi}}\ and\ \bibinfo {author} {\bibfnamefont {H.}~\bibnamefont
  {Umezawa}},\ }\bibfield  {title} {\bibinfo {title} {Thermo field dynamics},\
  }\href {https://doi.org/10.1142/s0217979296000817} {\bibfield  {journal}
  {\bibinfo  {journal} {International Journal of Modern Physics B}\ }\textbf
  {\bibinfo {volume} {10}},\ \bibinfo {pages} {1755} (\bibinfo {year}
  {1996})}\BibitemShut {NoStop}%
\bibitem [{\citenamefont {Chapman}\ \emph {et~al.}(2019)\citenamefont
  {Chapman}, \citenamefont {Eisert}, \citenamefont {Hackl}, \citenamefont
  {Heller}, \citenamefont {Jefferson}, \citenamefont {Marrochio},\ and\
  \citenamefont {Myers}}]{Chapman_SciPost_2019}%
  \BibitemOpen
  \bibfield  {author} {\bibinfo {author} {\bibfnamefont {S.}~\bibnamefont
  {Chapman}}, \bibinfo {author} {\bibfnamefont {J.}~\bibnamefont {Eisert}},
  \bibinfo {author} {\bibfnamefont {L.}~\bibnamefont {Hackl}}, \bibinfo
  {author} {\bibfnamefont {M.~P.}\ \bibnamefont {Heller}}, \bibinfo {author}
  {\bibfnamefont {R.}~\bibnamefont {Jefferson}}, \bibinfo {author}
  {\bibfnamefont {H.}~\bibnamefont {Marrochio}},\ and\ \bibinfo {author}
  {\bibfnamefont {R.~C.}\ \bibnamefont {Myers}},\ }\bibfield  {title} {\bibinfo
  {title} {Complexity and entanglement for thermofield double states},\ }\href
  {https://doi.org/10.21468%2Fscipostphys.6.3.034} {\bibfield  {journal}
  {\bibinfo  {journal} {{SciPost} Physics}\ }\textbf {\bibinfo {volume} {6}}
  (\bibinfo {year} {2019})}\BibitemShut {NoStop}%
\bibitem [{\citenamefont {Gregory}\ \emph {et~al.}(2023)\citenamefont
  {Gregory}, \citenamefont {Fernley}, \citenamefont {Tao}, \citenamefont
  {Bromley}, \citenamefont {Stepp}, \citenamefont {Zhang}, \citenamefont
  {Kotochigova}, \citenamefont {Hazzard},\ and\ \citenamefont
  {Cornish}}]{gregory2023secondscale}%
  \BibitemOpen
  \bibfield  {author} {\bibinfo {author} {\bibfnamefont {P.~D.}\ \bibnamefont
  {Gregory}}, \bibinfo {author} {\bibfnamefont {L.~M.}\ \bibnamefont
  {Fernley}}, \bibinfo {author} {\bibfnamefont {A.~L.}\ \bibnamefont {Tao}},
  \bibinfo {author} {\bibfnamefont {S.~L.}\ \bibnamefont {Bromley}}, \bibinfo
  {author} {\bibfnamefont {J.}~\bibnamefont {Stepp}}, \bibinfo {author}
  {\bibfnamefont {Z.}~\bibnamefont {Zhang}}, \bibinfo {author} {\bibfnamefont
  {S.}~\bibnamefont {Kotochigova}}, \bibinfo {author} {\bibfnamefont
  {K.~R.~A.}\ \bibnamefont {Hazzard}},\ and\ \bibinfo {author} {\bibfnamefont
  {S.~L.}\ \bibnamefont {Cornish}},\ }\href@noop {} {\bibinfo {title}
  {Second-scale rotational coherence and dipolar interactions in a gas of
  ultracold polar molecules}} (\bibinfo {year} {2023}),\ \Eprint
  {https://arxiv.org/abs/2306.02991} {arXiv:2306.02991 [physics.atom-ph]}
  \BibitemShut {NoStop}%
\bibitem [{\citenamefont {Christakis}\ \emph
  {et~al.}(2023{\natexlab{b}})\citenamefont {Christakis}, \citenamefont
  {Rosenberg}, \citenamefont {Raj}, \citenamefont {Chi}, \citenamefont
  {Morningstar}, \citenamefont {Huse}, \citenamefont {Yan},\ and\ \citenamefont
  {Bakr}}]{christakis2023probing}%
  \BibitemOpen
  \bibfield  {author} {\bibinfo {author} {\bibfnamefont {L.}~\bibnamefont
  {Christakis}}, \bibinfo {author} {\bibfnamefont {J.~S.}\ \bibnamefont
  {Rosenberg}}, \bibinfo {author} {\bibfnamefont {R.}~\bibnamefont {Raj}},
  \bibinfo {author} {\bibfnamefont {S.}~\bibnamefont {Chi}}, \bibinfo {author}
  {\bibfnamefont {A.}~\bibnamefont {Morningstar}}, \bibinfo {author}
  {\bibfnamefont {D.~A.}\ \bibnamefont {Huse}}, \bibinfo {author}
  {\bibfnamefont {Z.~Z.}\ \bibnamefont {Yan}},\ and\ \bibinfo {author}
  {\bibfnamefont {W.~S.}\ \bibnamefont {Bakr}},\ }\bibfield  {title} {\bibinfo
  {title} {Probing site-resolved correlations in a spin system of ultracold
  molecules},\ }\href {https://www.nature.com/articles/s41586-022-05558-4}
  {\bibfield  {journal} {\bibinfo  {journal} {Nature}\ }\textbf {\bibinfo
  {volume} {614}},\ \bibinfo {pages} {64} (\bibinfo {year}
  {2023}{\natexlab{b}})}\BibitemShut {NoStop}%
\bibitem [{\citenamefont {Bornet}\ \emph {et~al.}(2024)\citenamefont {Bornet},
  \citenamefont {Emperauger}, \citenamefont {Chen}, \citenamefont {Machado},
  \citenamefont {Chern}, \citenamefont {Leclerc}, \citenamefont
  {G{\ifmmode\acute{e}\else\'{e}\fi}ly}, \citenamefont {Barredo}, \citenamefont
  {Lahaye}, \citenamefont {Yao},\ and\ \citenamefont
  {Browaeys}}]{bornet2024enhancing}%
  \BibitemOpen
  \bibfield  {author} {\bibinfo {author} {\bibfnamefont {G.}~\bibnamefont
  {Bornet}}, \bibinfo {author} {\bibfnamefont {G.}~\bibnamefont {Emperauger}},
  \bibinfo {author} {\bibfnamefont {C.}~\bibnamefont {Chen}}, \bibinfo {author}
  {\bibfnamefont {F.}~\bibnamefont {Machado}}, \bibinfo {author} {\bibfnamefont
  {S.}~\bibnamefont {Chern}}, \bibinfo {author} {\bibfnamefont
  {L.}~\bibnamefont {Leclerc}}, \bibinfo {author} {\bibfnamefont
  {B.}~\bibnamefont {G{\ifmmode\acute{e}\else\'{e}\fi}ly}}, \bibinfo {author}
  {\bibfnamefont {D.}~\bibnamefont {Barredo}}, \bibinfo {author} {\bibfnamefont
  {T.}~\bibnamefont {Lahaye}}, \bibinfo {author} {\bibfnamefont {N.~Y.}\
  \bibnamefont {Yao}},\ and\ \bibinfo {author} {\bibfnamefont {A.}~\bibnamefont
  {Browaeys}},\ }\bibfield  {title} {\bibinfo {title} {{Enhancing a Many-body
  Dipolar Rydberg Tweezer Array with Arbitrary Local Controls}},\ }\bibfield
  {journal} {\bibinfo  {journal} {arXiv}\ }\href
  {https://doi.org/10.48550/arXiv.2402.11056} {10.48550/arXiv.2402.11056}
  (\bibinfo {year} {2024}),\ \Eprint {https://arxiv.org/abs/2402.11056}
  {2402.11056} \BibitemShut {NoStop}%
\end{thebibliography}%



\cleardoublepage
\appendix

\setcounter{equation}{0}
\setcounter{figure}{0}
\setcounter{table}{0}
\makeatletter
\renewcommand{\theequation}{S\arabic{equation}}
\renewcommand{\thefigure}{S\arabic{figure}}

%
\setcounter{equation}{0}
\setcounter{figure}{0}
\setcounter{table}{0}
\makeatletter
\renewcommand{\thefigure}{S\arabic{figure}}
\section{Supplementary Information}
This supplementary information contains additional details on the Bogoliubov analysis, the Chebyshev method, the equilibrium distribution, the entanglement entropy, and extended numerical results.


\subsection{Bogoliubov Analysis}
In this section, we provide further details on the Bogoliubov analysis.
We write the Hamiltonian of the two-chain system in the quasi-momentum representation
\begin{equation}
\begin{split}
\hat{H} &= \sum_{k} \varepsilon_{k}(\hat{a}_{k}^{\dagger}\hat{a}_{k}+\hat{b}_{k}^{\dagger}\hat{b}_{k}) +\\
&\sum_{k}[|\Omega_{k}|e^{-i\alpha_{k}}\hat{a}_{k}^{\dagger}\hat{b}_{-k}^{\dagger}+|\Omega_{k}|e^{i\alpha_{k}}\hat{b}_{-k}\hat{a}_{k}], 
\end{split}
\label{BGT1}
\end{equation}
where we have made explicit the complex nature of the inter-chain coupling $\Omega_{k} = |\Omega_{k}|e^{-i\alpha_{k}}$. 
The Hamiltonian can be diagonalized by means of a Bogoliubov transformation
\begin{equation}
\begin{split}
\hat{\gamma}_{k} &= u_{k}\hat{a}_{k}-v_{k}\hat{b}_{-k}^{\dagger},\\
\hat{\beta}_{-k}^{\dagger} &= u_{k}^{*}\hat{b}_{-k}^{\dagger}-v_{k}^{*}\hat{a}_{k},
\label{BGT2}
\end{split}
\end{equation}
where $\hat{\gamma}_{k}$ and $\hat{\beta}_{-k}^{\dagger}$ are the bosonic operators associated with the Bogoliubov excitations, and $u_{k}$ and $v_{k}$ obey the Bogoliubov-de-Gennes equations
\begin{equation}
\begin{split}
\xi_{k}u_{k} &= \varepsilon_{k}u_{k}+|\Omega_{k}|e^{-i\alpha_{k}}v_{k}^{*},\\
\xi_{k}v_{k} &=-|\Omega_{k}|e^{i\alpha_{k}}u_{k}-\varepsilon_{k}v_{k}^{*},
\end{split}
\label{BGT3}
\end{equation}
where $\xi_{k} = \sqrt{\varepsilon_{k}^{2}-|\Omega_{k}|^2}$ is the Bogoliubov spectrum. Real and imaginary Bogoliubov eigenenergies give rise to different excitation dynamics. In the case of real eigenvalues, the time dependence of the Bogoliubov operators is $\hat{\gamma}_{k}(t) = e^{-i\xi_{k}t}\hat{\gamma}_{k}(0)$ and $\hat{\beta}_{-k}^{\dagger}(t) = e^{i\xi_{k}t}\hat{\beta}_{-k}^{\dagger}(0)$.  By inverting the Bogoliubov transformation, one can obtain
the time-dependence of the bosonic operators
\begin{equation}
\begin{split}
\hat{a}_{k}(t) &=\hat{a}_{k}\left(\cos\xi_{k}t-\frac{i\varepsilon_{k}}{\xi_{k}}\sin\xi_{k}t\right)\\
&-i\hat{b}_{-k}^{\dagger}\left(\frac{e^{i\alpha_{k}}|\Omega_{k}|}{\xi_{k}}\right)\sin\xi_{k}t,\\
\hat{b}_{-k}^{\dagger}(t) &=\hat{b}_{-k}^{\dagger}\left(\cos\xi_{k}t+\frac{i\varepsilon_{k}}{\xi_{k}}\sin\xi_{k}t\right)\\
&+i\hat{a}_{k}\left(\frac{e^{-i\alpha_{k}}|\Omega_{k}|}{\xi_{k}}\right)\sin\xi_{k}t.
\end{split}
\label{BGT4}
\end{equation}
With the aid of the above equations, the vacuum expectation value of the population of spin excitations in mode $k$ of the leg A is $n_{k}^{A}(t) = \langle 0|(\hat{a}_{k}^{\dagger}\hat{a}_{k})(t)|0\rangle=[|\Omega_{k}|\sin(\xi_{k}t)/\xi_{k}]^{2}$. Integration of $n_{k}^{A}(t)$ over the first Brillouin zone gives the total density of excitations
\begin{equation}
n^{A}(t) = \int_{BZ}\frac{dk}{2\pi a} \ |\Omega_{k}|^{2}\left[\frac{\sin|\xi_{k}|t}{|\xi_{k}|}\right]^{2}.
\label{BGT5}
\end{equation}
For Bogoliubov modes with imaginary eigenenergies, the time-dependence of the bosonic operators is 
\begin{equation}
\begin{split}
\hat{a}_{k}(t) &=\hat{a}_{k}\left(\cosh|\xi_{k}|t-\frac{i\varepsilon_{k}}{|\xi_{k}|}\sinh|\xi_{k}|t\right),\\
&-i\hat{b}_{-k}^{\dagger}\left(\frac{e^{i\alpha_{k}}|\Omega_{k}|}{|\xi_{k}|}\right)\sinh|\xi_{k}|t\\
\hat{b}_{-k}^{\dagger}(t) &=\hat{b}_{-k}^{\dagger}\left(\cosh|\xi_{k}|t+\frac{i\varepsilon_{k}}{|\xi_{k}|}\sin|\xi_{k}|t\right)\\
&+i\hat{a}_{k}\left(\frac{e^{-i\alpha_{k}}|\Omega_{k}|}{|\xi_{k}|}\right)\sin|\xi_{k}|t.
\end{split}
\label{BGT6}
\end{equation}
As mentioned in the main text, imaginary eigenvalues yield exponential growth of spin excitations, that is $n_{k}^{A}(t) = [|\Omega_{k}|\sinh(|\xi_{k}|t)/|\xi_{k}|]^{2}\propto(|\Omega_{k}|/|\xi_{k}|)^{2}e^{2|\xi_{k}|t}=(|\Omega_{k}|/|\xi_{k}|)^{2}e^{2\Gamma_{k}t}$, where $\Gamma_{k}$ is the rate of the exponential growth. Since $\sin(i|\xi_{k_{c}}|)/i|\xi_{k_{c}}|\rightarrow -\sinh(|\xi_{k_{c}}|)/|\xi_{k_{c}}|$, one can safely use the expression in Eq. (\ref{BGT5}) to obtain the time dependence of the density of excitations in each leg. 


\section{Chebyshev expansion}
\label{Chebyshev}
In the following, we briefly introduce the Chebyshev algorithm used to calculate the time evolution of the spin ladder. A comprehensive description of the method can be found in Ref.~\cite{fehske2007computational}. Considering that at time $t=0$ a given quantum system is prepared in the state $|\psi_{0}\rangle$, its state at a later time is given by $|\psi_{t}\rangle=e^{-i\hat{H}t}|\psi_{0}\rangle$ ($\hbar=1$). Thus, by calculating an approximate form of the time evolution operator $\hat{U}(t) = e^{-i\hat{H}t}$, one can estimate the evolution of the quantum system. Such an approximation is given by the Chebyshev expansion
\begin{equation}
U(t) = e^{-i(a\tilde{H}+b)t} \approx e^{-ibt}\left(c_{0} + 2\sum_{\nu=1}^{N}c_{\nu}T_{\nu}(\tilde{H})\right),
\label{}    
\end{equation}
where $a=(E_{max}-E_{min})/2$, $b=(E_{max}+E_{min})/2$ with
$E_{max}$ ($E_{min}$) the highest (lowest) eigenstate energy of the Hamiltonian $\hat{H}$, $\tilde{H} = (\hat{H}-b)/a$ is a rescaling of the original Hamiltonian, and $T_{\nu}(x)$ is the Chebyshev polynomial of order $\nu$. The coefficients $c_{\nu}$ of the Chebyshev expansion are given by $c_{\nu} = (-i)^{\nu}J_{\nu}(at)$, where $J_{\nu}(x)$ denotes the Bessel function of order $\nu$. The order $N$ of the Chebyshev expansion is fixed by demanding a unitary time evolution. 


\section{Equilibrium distribution}
In this section, we provide further details concerning the equilibrium distribution that characterizes the ergodic regime.
Following Ref.~\cite{lepoutre2019out}, the long-time behavior of local observables, such as the leg magnetization $m_{A}=\sum_{i=1}^{L}\hat{s}_{i,A}$, can be described by a thermal distribution of the form
\begin{equation}
\hat{\rho}_{th} = \frac{1}{Z}\text{exp}\left(-\beta\hat{H}-\mu_{1}\hat{S}_{z}-\mu_{2}(\hat{S}_{z}^{2}-\langle\hat{S}_{z}\rangle^{2})\right),
\label{AQTeq1}
\end{equation}
where
\begin{equation}
Z = \text{Tr}\left[\text{exp}\left(-\beta\hat{H}-\mu_{1}\hat{S}_{z}-\mu_{2}(\hat{S}_{z}^{2}-\langle\hat{S}_{z}\rangle^{2})\right)\right]    
\label{AQTeq2}
\end{equation}
is the partition function, $\beta$, $\mu_{1}$, and $\mu_{2}$ are the Lagrange multipliers associated with the inverse temperature, total magnetization, and the variance of the magnetization, respectively. Since in the two-chain system spin flips are created in pairs, one per chain, the matrix elements of $\hat{S}_{z}$ and $\hat{S}_{z}^{2}$ are zero. Furthermore, it is straightforward to notice that the diagonal matrix elements of $\hat{H}$ are also zero. Therefore, the density matrix is proportional to the identity operator $\hat{I}$ and the system can be described using the microcanonical ensemble, where each microstate has equal probability. In this scenario, the partition function is simply given by the Hilbert space dimension
\begin{equation}
Z = \binom{2L}{L}=\frac{(2L)!}{L!^{2}}.    
\label{SI_AQTeq3}
\end{equation}
The corresponding density matrix is $\hat{\rho}_{th}=\frac{\hat{I}}{Z}$. 
Tracing over one chain we obtain the reduced density matrix. Since the matrix elements of $\hat{\rho}^{th}$ are all equal, this is a straightforward task, resulting in 
\begin{equation}
\rho^{th}_{A}(N_A) = \frac{\binom{L}{N_A}^{2}}{\binom{2L}{L}}.    
\label{SI_AQTeq4}
\end{equation}
Using the Stirling approximation for large $L$ results in the Gaussian distribution discussed in the main text.



\begin{figure}[t!]
\centering
\includegraphics[width=1.0\columnwidth]{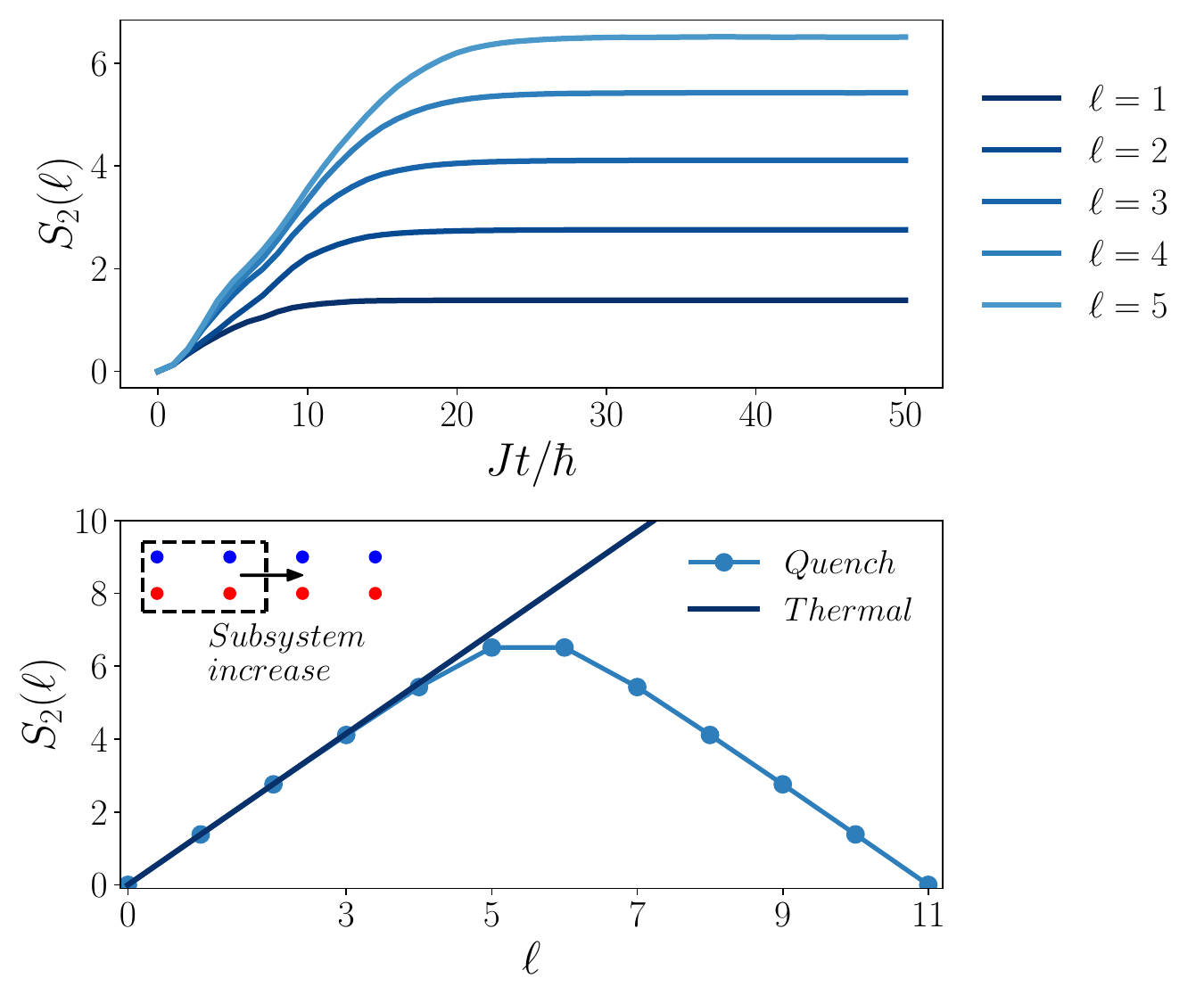}
\caption{(a) R\'enyi entropy as a function of time for several subsystem sizes. (b) Saturated value of the R\'enyi entropy as a function of the subsystem size. The dark blue line represents the value of $S_{2}(\ell)$ obtained from a thermal density matrix. The inset shows the subsystem increase considered. In both panels we consider $(\theta,\Delta)=(0,2)$ where the system is in the ergodic regime. Results obtained from Chebyshev calculations for $11$ rungs.}
\label{FigEx2-1}
\end{figure}




\begin{figure}[h!]
\centering
\includegraphics[width=1.0\columnwidth]{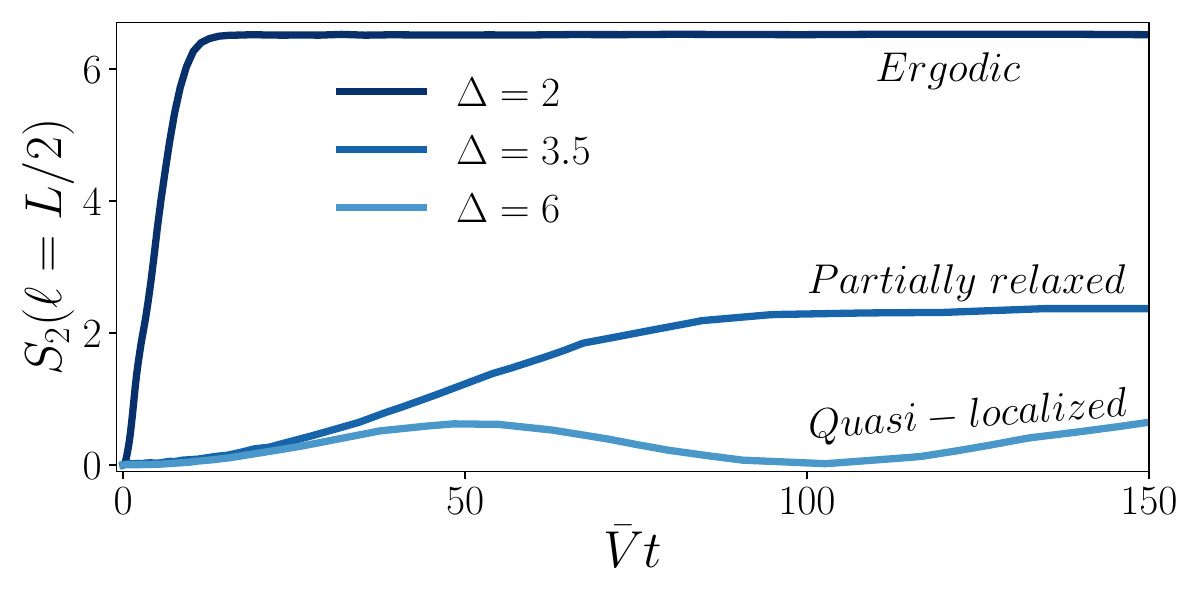}
\caption{Evolution of the half-system R\'enyi entropy in the three different relaxation regimes. We consider $\theta=0$. Results obtained from Chebyshev calculations for $11$ rungs.}
\label{FigEx2-2}
\end{figure}



\section{Entanglement entropy}
We briefly discuss at this point the dynamics of the entanglement entropy. In particular, we focus on the second-order R\'enyi entropy $S_{2}(\ell) = -\log \text{Tr}({\hat{\rho}_{\ell}}^{2})$, where ${\hat{\rho}_{\ell}}$ is density matrix associated with a subsystem of size $\ell$ (see inset of Fig. \ref{FigEx2-1}(b)). In Fig. \ref{FigEx2-1}(a), we plot the time evolution of $S_{2}(\ell)$ for different values of $\ell$. At short times, the entanglement entropy shows an approximately linear increase, regardless of the size of the subsystem. Subsequently, the dynamics becomes subsystem-size dependent. Eventually, in the long-time limit, the entanglement entropy reaches a steady-state value consistent with the corresponding entropy of a thermal density matrix (see Fig. \ref{FigEx2-1}(b)). Notice that as $\ell$ becomes comparable to the full system size, the subsystem R\'enyi entropy bends back to zero, indicating that the entire system is in a pure state.

To conclude this section, in Fig. \ref{FigEx2-2}, we illustrate the half-system R\'enyi entropy as a function of time in the three different relaxation regimes. As one can notice, for $\Delta=2$ (ergodic), $S_{2}$ rapidly increases to its maximum value, indicating thermalization. In the partially-relaxed regime, the R\'enyi entropy initially increases until it reaches a saturation point, which deviates from the thermal equilibrium value. Subsequently, it stabilizes at this constant value before eventually rising to the anticipated equilibrium value at extremely long times. Within the quasi-localized regime, the entanglement entropy exhibits long-term oscillations around a small value, suggesting that the system experiences minimal departure from its initial state.


\begin{figure}[t!]
\centering
\includegraphics[width=1.0\columnwidth]{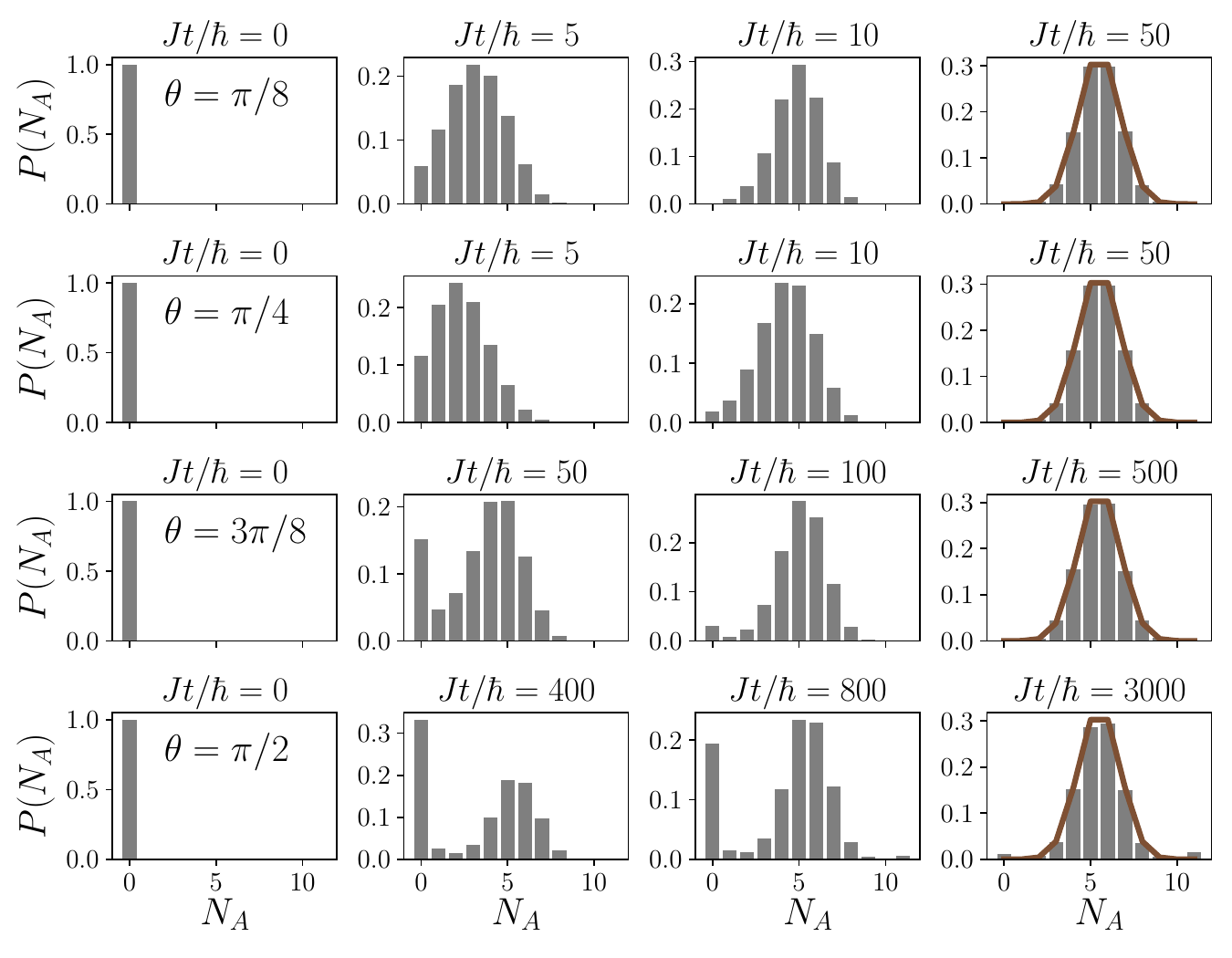}
\caption{Evolution of the distribution of spin excitations in chain A for different dipole orientations. The brown curve depicts the expected thermal distribution. Note the vastly longer time scales in the last two rows. Results obtained from Chebyshev calculations for $11$ rungs and $\Delta = 2$.} 
\label{FigEx1}
\end{figure}



\begin{figure}[h!]
\centering
\includegraphics[width=1.0\columnwidth]{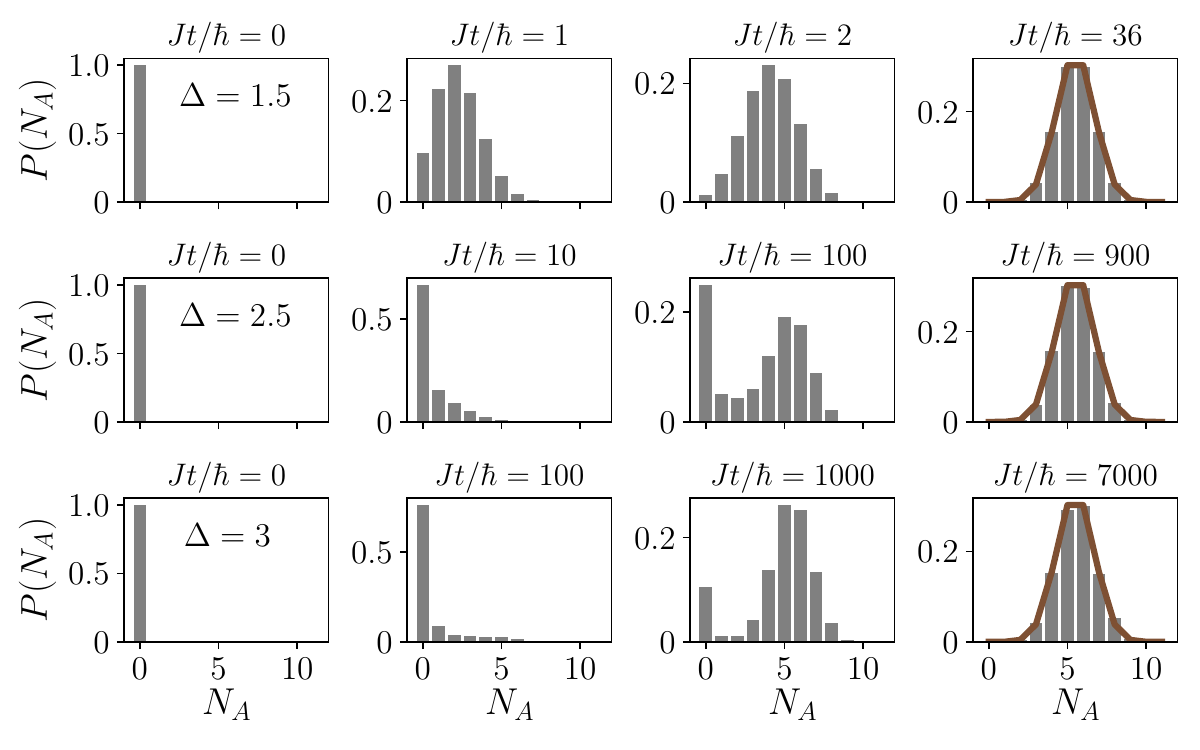}
\caption{Evolution towards thermalization of the distribution of spin excitations in chain A for $\theta=0$ and different separations of the chains. The brown curve is associated with the expected thermal distribution. Note again the much longer time scales in the last two rows. Results obtained from Chebyshev calculations for $11$ rungs.}
\label{FigEx2}
\end{figure}



\section{Extended results}
In this section, we provide extended results for the time dependence of the distribution of spin excitations. In Fig. \ref{FigEx1}, we show the evolution of the distribution of spin excitations in chain A for different dipole orientations and fixed separation of the chains $\Delta=2$. As one can readily observe, regardless of the angle $\theta$, the distribution of excitations attains an equilibrium state consistent with the thermal prediction. It is important to note the role of the dipole orientation in the time scale to reach the thermal state, while for $\theta=\pi/8$ and $\pi/4$ excitations are created rapidly, bringing the system to a homogeneous state, for $\theta=3\pi/8$ and $\pi/2$ there is considerable overlap with the initial state up to very long time scales.

To show the role of the separation of the chains on the dynamics of the distribution of excitations, in Fig. \ref{FigEx2}, we plot $\rho_{A}(n)$ for three different values of $\Delta$. As one can notice, thermalization is reached for the three different values of the separations of the chains considered, however, the time scale required to enter such steady-state increases enormously.  Note that cases well inside the partially-relaxed regime, e.g. the lower panels of Fig.~\ref{FigEx1} and~\ref{FigEx2}, keep the bimodal distribution for very long times. The eventual relaxation 
towards the Gaussian distribution demands evolution times typically well above the longest rotational coherent times realized in experiments. 

\begin{figure}[h!]
\centering
\includegraphics[width=0.85\columnwidth]{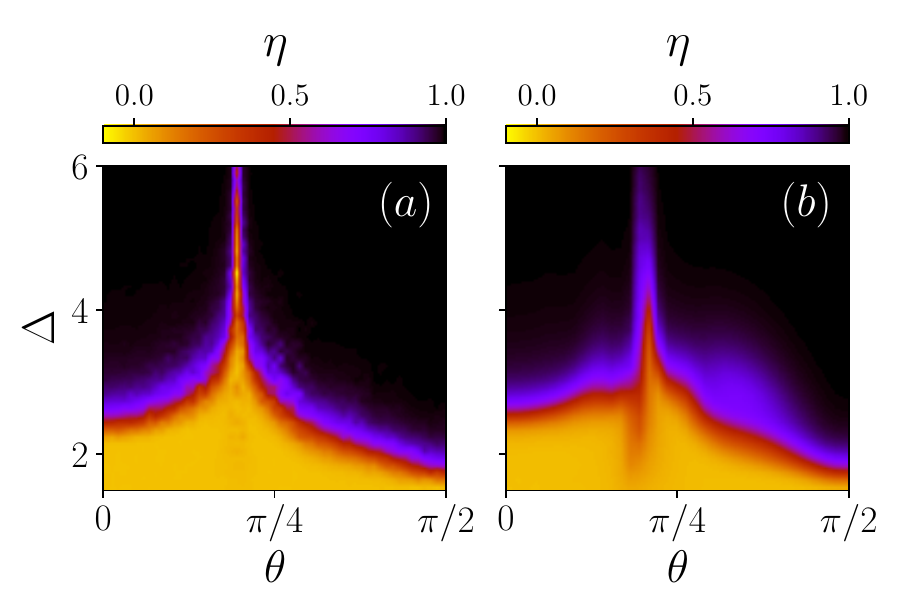}
\caption{Imbalance $\eta$ after a time $Jt/\hbar=50$ as a function of $\theta$ and $\Delta$ for a fully-filled lattice with a random on-site magnetic field $W/J=0.15$ (a) and $W/J=0.3$ (b), respectively. Results obtained from Chebyshev calculations for $11$ rungs and $100$ disorder realizations.}
\label{FigEx3}
\end{figure}

Imperfections of the lattice or the tweezer array may result in a random effective local magnetic field, $h_{j,\alpha} \hat s^z_{j\alpha}$. In Fig.~\ref{FigEx3}~(a)-\ref{FigEx3}~(b) we consider the effect of uniformly distributed random $-W/2<h_{j,\alpha}<W/2$, with $W/J=0.15$ and with $W/J=0.3$, respectively. As pointed out in the manuscript, the random disorder only slightly affects the imbalance behavior.


\end{document}